\documentclass[useAMS,usenatbib,usegraphicx,letterpaper]{mn2e}

\usepackage{multirow} 
\usepackage{epsfig}
\usepackage{subfigure}

\newcommand{\etal}{et al.}

\def\mnras{MNRAS}
\def\nat{Nature}
\def\apj{ApJ}
\def\apjs{ApJS}
\def\apjl{ApJ}
\def\aap{A\&A}
\def\aj{AJ}
\def\pasj{PASJ}
\def\procspie{Proc. SPIE}
\def\pasp{Proc. of the Astronomical Society of the Pacific}

\title[Monitoring of the X--ray weak quasar PHL~1092] {Insights on the
  X--ray weak quasar phenomenon from {\it XMM--Newton} monitoring of PHL~1092} 
\author[G.\ Miniutti \etal]
{\parbox{\textwidth}{G.~Miniutti,$^{1}$\thanks{E-mail: \texttt{gminiutti@cab.inta-csic.es}}
W.~N.~Brandt,$^{2,3}$
D.~P.~Schneider,$^{2,3}$
A.~C.~Fabian,$^{4}$
L.~C.~Gallo,$^{5}$ and 
Th.~Boller $^{6}$}\vspace{0.5cm}\\
\parbox{\textwidth}{
$^{1}$Centro de Astrobiolog\'{i}a (CSIC--INTA), Dep. de Astrof\'{i}sica; 
ESA, P.O: Box 78, E-28691, Villanueva de la Ca\~nada, Madrid, Spain\\
$^{2}$Department of Astronomy and Astrophysics, The Pennsylvania State 
University, 525 Davey Lab., Univeristy Park, PA 16802, USA\\
$^{3}$Institute for Gravitation and the Cosmos, The Pennsylvania State 
University, University Park, PA 16802, USA\\
$^{4}$Institute of Astronomy, Madingley Road, Cambridge CB3 0HA\\
$^{5}$ Department of Astronomy \& Physics, Saint Mary's University, 
923 Robie Street, Halifax, NS B3H 3C3\\
$^{6}$Max--Planck--Institut f\"ur extraterrestrische Physik, Postfach 
1312, 85741 Garching, Germany}}

\pagerange{\pageref{firstpage}--\pageref{lastpage}}
\pubyear{2012}

\usepackage{times}

\begin{document}

\label{firstpage}

\maketitle

\begin{abstract}
PHL~1092 is a $z\sim 0.4$ high--luminosity counterpart of the class of
Narrow--Line Seyfert~1 galaxies. In 2008, PHL~1092 was found to be in
a remarkably low X--ray flux state during an {\it XMM--Newton}
observation. Its 2~keV flux density had dropped by a factor of
$\sim$260 with respect to a previous observation performed 4.5~yr
earlier. The UV flux remained almost constant, resulting in a
significant steepening of the optical--to--X--ray slope
$\alpha_{\rm{ox}}$ from $-1.57$ to $-2.51$, making PHL~1092 one of the
most extreme X--ray weak quasars with no observed broad absorption
lines (BALs) in the UV. We have monitored the source since 2008 with
three further {\it XMM--Newton} observations, producing a simultaneous
UV and X--ray database spanning almost 10~yr in total in the activity
of the source. Our monitoring program demonstrates that the
$\alpha_{\rm{ox}}$ variability in PHL~1092 is entirely driven by
long--term X--ray flux changes. We apply a series of
physically--motivated models with the goal of explaining the
UV--to--X--ray spectral energy distribution (SED) and the extreme
X--ray and $\alpha_{\rm{ox}}$ variability. We consider three possible
models: i) A {\it breathing corona} scenario in which the size of the
X--ray emitting corona is correlated with the X--ray flux. In this
case, the lowest X--ray flux states of PHL~1092 are associated with an
almost complete collapse of the X--ray corona down to the marginal
stable orbit; ii) An absorption scenario in which the X--ray flux
variability is entirely due to intervening absorption. If so, PHL~1092
is a quasar with standard X--ray output for its optical luminosity,
appearing as X--ray weak at times due to absorption; iii) A
disc--reflection--dominated scenario in which the X--ray emitting
corona is confined within a few gravitational radii from the black
hole at all times. In this case, the intrinsic variability of PHL~1092
only needs to be a factor of $\sim 10$ rather than the observed factor
of $\sim 260$. We discuss these scenarios in the context of non--BAL
X--ray weak quasars.
\end{abstract}


\section{Introduction}
\label{intro}

X--ray emission is a universal property of efficiently accreting black
holes (BH) and its origin is, most likely, Compton up--scattering of
the UV/EUV disc emission in an X--ray corona of hot electrons
(e.g. Haardt, Maraschi \& Ghisellini 1994). There are, however, a few
examples of Active Galactic Nuclei (AGN) where the corona appears to
emit X--rays much less efficiently than usual, with an X--ray output
$\sim$~10--100 times lower than that expected based upon the
optical/UV emission properties. These are known as X--ray weak AGN
which often exhibit broad absorption lines (BALs) in the UV,
suggesting an absorption--driven interpretation of their X--ray
weakness (Brandt, Laor \& Wills 2000). Indeed, X--ray absorption has
been convincingly detected in a number of BAL quasars (e.g. Brinkmann
et al. 1999; Wang et al. 1999; Gallagher et al. 2001, 2004; Mathur et
al. 2001; Chartas et al. 2002, 2007, 2009).

However, a growing number of X--ray weak quasars do not show any
detectable BALs in their UV spectra (e.g. Wu et al. 2011). The
best--studied case is that of PHL~1811 (Leighly et al. 2001, 2007a,
2007b). This quasar has been observed several times in X--rays and is
persistently X--ray weak. Its X--ray spectrum shows no sign of
absorption of a higher level X--ray flux, so that PHL~1811 appears to
be intrinsically X--ray weak. The C~\textsc{iv} emission line is weak,
blueshifted and asymmetric, indicating the presence of a wind. The
near--UV spectrum is dominated by Fe~\textsc{ii} and Fe~\textsc{iii}
emission--line blends and unusually low--ionisation lines. Leighly et
al. (2007b) demonstrated, using detailed photo--ionisation modelling, that
the optical/UV emission lines are consistent with a wind plus disc
scenario in which C~\textsc{iv} is produced in the outflow while the
low--ionisation lines (e.g. Mg~\textsc{ii}) are associated with the
denser disc at larger radii and see an ionising continuum filtered
through the wind itself. Indeed, a soft, X--ray weak ionising
continuum can explain the unusual optical/UV line properties if
filtering through the wind is considered for the disc--lines. The
question in PHL~1811 and similar sources is then whether the X--ray
emission is intrinsically weak or due to intervening X--ray--only
absorption.

Here we consider the case of PHL~1092 which recently became one of the
most extreme X--ray weak non--BAL quasars known (see below). PHL~1092
($z=0.396$) is a radio--quiet quasar with outstanding Fe~\textsc{ii}
emission (Bergeron \& Kunth 1980, 1984; Kwan et al 1995). Its broad
line widths of $\sim$~1800~km~s$^{-1}$ and a ratio
[O~\textsc{iii}]$\lambda$5007~/~H$\beta \sim 0.9$, together with the
strong Fe~\textsc{ii} emission (Fe~\textsc{ii}/H$\beta =1.81$),
classify PHL~1092 as a high--luminosity Narrow--Line Seyfert~1 (NLS1)
galaxy (Osterbrock \& Pogge 1985).  The bolometric luminosity of
PHL~1092 is $4-5\times 10^{46}$~erg~s$^{-1}$. Its UV spectrum shows
relatively weak, broad, and blueshifted C~\textsc{iv} emission and
presents many analogies with the prototypical non--BAL X--ray weak
quasar PHL~1811 (Wu et al. 2011).

In 2008, PHL~1092 was found to be in a remarkably low X--ray flux
state during an {\it XMM--Newton} observation with a 2~keV
flux--density drop of $\sim$~260 with respect to a previous 2003 {\it
  XMM--Newton} exposure (Miniutti et al. 2009b). The UV flux
remained almost constant, resulting in a significant steepening of
the optical--to--X--ray slope $\alpha_{\rm{ox}}$ from $-1.57$ to
$-2.51$ and showing that even extreme X--ray weakness can in fact be a
transient phenomenon. Quasi--simultaneous optical spectra from
Mg~\textsc{ii} to H$\beta$ did not reveal significant changes with
respect to historical ones allowing us to exclude the presence of a
Mg~\textsc{ii} BAL.

In order to shed light on the X--ray weakness phenomenon in non--BAL
quasars, we have monitored PHL~1092, obtaining three 
observations with {\it XMM--Newton} since 2008, and another optical
spectrum which is quasi--simultaneous with our last X--ray pointing
in 2010. We report here results from our monitoring campaign spanning
almost 10~yr in total in the activity of the source. Our observations
provide simultaneous UV and X--ray data with which we follow the
long--term X--ray and optical/UV variability of this remarkable
source. We consider a series of physically motivated models that can
reproduce the UV--to--X--ray data at the different probed X--ray flux
levels, and we discuss the implications of our results in the context
of non--BAL X--ray weak quasars. 

\section{Observations}

PHL~1092 was observed with {\it XMM--Newton} on 6 occasions. The first
observation was performed on 2000 July 31, but due to problems with
the European Photon Imaging Camera (EPIC) cameras (Str{\"u}der et
al. 2001; Turner et al. 2001), only X--ray and UV light curves could
be obtained. The source was re--observed three years later, resulting
in a complete set of ODF files for all detectors. Results from both
observations were reported by Gallo et al. (2004). We obtained a
further $\sim$~60~ks observation on 2008 January 20 which caught the
source in an extreme X--ray weak state with a soft X--ray flux drop of
$\sim$260 with respect to the 2003 observation as discussed by
Miniutti et al. (2009b). The source was monitored during the two
subsequent years, with three {\it XMM--Newton} observations performed
on 2009 June 27, 2010 July 7, and 2010 December 30. The 2008
observation was complemented with quasi--simultaneous optical spectra
at the Hobby--Eberly and William Herschel Telescopes on Januar 30 and
February 14 respectively, and an additional optical spectrum was
obtained at the Hobby--Eberly Telescope on 2010 October 31. The HET
data cover the $\approx$~4300--7300~\AA\ spectral range, while the WHT
spectrum is obtained in the $\approx$~3600--5180~\AA\ and
$\approx$~5950--9090~\AA range.

The observation log of the {\it XMM--Newton} observations is given in
Table~\ref{tab1}. The {\it XMM--Newton} ODF files were processed using
the dedicated Science Analysis Software ({\footnotesize{SAS v11.0}})
to obtain calibrated event lists which were filtered for bad pixels
and high background flaring. Single and double events were selected to
extract scientific products for the EPIC detectors. We optimised the
spectral extraction areas of the source to maximise the
signal--to--noise ratio resulting in observation--dependent circular
extraction regions of radii in the range of 25$''$-40$''$. Background
spectra were extracted from larger off--source areas. Here we make use
of the EPIC--pn data from all observations because of its higher
sensitivity with respect to the MOS detectors. However, we have
checked that the spectra obtained from the pn and MOS detectors are
consistent with each other, which is indeed the case. All pn spectra
were grouped to produce a minimum of 25 counts per energy channel. We
also report results from the {\it XMM--Newton} Optical Monitor (OM),
which was used to derive the monochromatic fluxes in the different
available filters in each observation. The OM generally takes multiple
images per filter. As the intra--observation variability was found to
be marginal at best, the fluxes reported in Table~\ref{tab1} are
obtained from the average count rate in each observation and filter using the
conversion factors between OM count rate and monochromatic flux
reported in the {\it XMM--Newton} SAS user
guide.\footnote{http://xmm.esa.int/sas/current/howtousesas.shtml}
Throughout this paper, we adopt a cosmology with ${\rm{H_0 =
    70~km~s^{-1}~Mpc^{-1}}}$, ${\rm{\Omega_\Lambda = 0.73}}$, and
${\rm{\Omega_M = 0.27}}$.

\section{Dramatic long--term X--ray variability}

\begin{figure}
\begin{center}
\includegraphics[width=0.33\textwidth,height=0.45\textwidth,angle=-90]{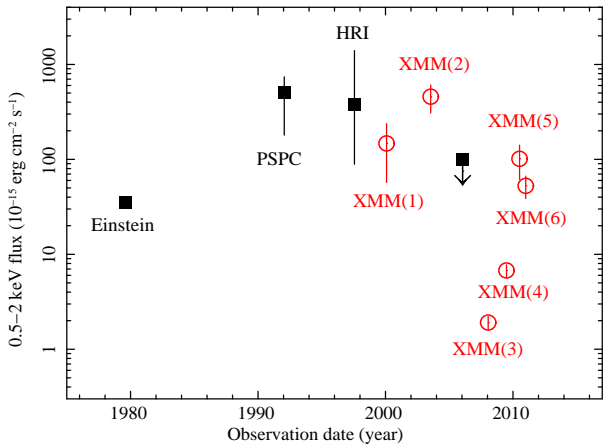}
\caption{The historical 0.5--2~keV X--ray flux of PHL~1092. Open
  symbols represent the {\it XMM--Newton} observations discussed in
  our paper. The data points represent the mean flux level during
  the observation, and the error is a measure of the
  intra--observation variability rather than of the statistical error on
  the mean flux. The upper limit is from an {\it XMM--Newton} slew in 2006. The
  Figure is an update of Fig~2 in Miniutti et al. (2009b). We refer to
  the caption of that Figure for further details.}
\label{xlc}
\end{center}
\end{figure}

The historical soft X--ray flux light curve of PHL~1092 is shown in
Fig.~\ref{xlc}. As can be seen, PHL~1092 was observed in an extremely
low flux state in 2008, which was reported and discussed by Miniutti et
al. (2009b). Its soft X--ray flux then recovered to reach in 2010
almost the same flux level it had in 2000. The earliest available
X--ray observation ({\it Einstein}, Wilkes et al. 1994) caught the
source in a relatively low--flux state in 1979. {\it ROSAT} observed
the source twice (PSPC: Forster \& Halpern 1996; Lawrence et al. 1997;
HRI: Brandt et al. 1999) at flux levels comparable to the highest
fluxes observed with {\it XMM--Newton}. A further {\it ASCA}
observation is not shown here as it was simultaneous with the {\it
  ROSAT} HRI monitoring campaign (Leighly et al 1999a and 1999b).

The {\it XMM-Newton} observations provide simultaneous photometric
data in the UV via the OM. The X--ray and UV monochromatic fluxes from
all {\it XMM-Newton} observations are reported in
Table~\ref{tab1}. The monochromatic 2~keV (rest--frame) flux varies by
a factor $\sim$~260 between observations XMM(2) and XMM(3). The UV
variations, however, have much smaller amplitude and they are not
strongly correlated with the X--ray ones. In fact, the maximum UV
variation is a $\sim$~15\% drop of the 1519~\AA\ (rest--frame)
monochromatic flux between observations XMM(3) and XMM(4), accompanied
by an X--ray flux increase by a factor $\sim$~4.2. The presence of
simultaneous UV and X--ray data allows us to derive reliably the
optical--to--X--ray slope $\alpha_{\rm{ox}}$ from 2000 to 2010. Here
we use the standard definition of $\alpha_{\rm{ox}} =
\log(f_{\rm{2}}/f_{\rm{2500}})~\log(\nu_{\rm{2}}/\nu_{\rm{2500}})^{-1}
= 0.384~\log(f_{\rm{2}}/f_{\rm{2500}})$ between 2~keV and 2500~\AA\ in
the rest--frame. Following Gibson, Brandt \& Schneider (2008), we also
define $\Delta\alpha_{\rm{ox}}$ as the difference between the observed
$\alpha_{\rm{ox}}$ and that expected from the anti--correlation
between $\alpha_{\rm{ox}}$ and the 2500~\AA\ monochromatic luminosity
which implies $\alpha_{\rm{ox}}^{\rm{expected}} = -1.48$ for PHL~1092
(e.g. Just et al. 2007).The $\alpha_{\rm{ox}}$ and
$\Delta\alpha_{\rm{ox}}$ values for each {\it XMM--Newton} observation
of PHL~1092 are reported in Table~\ref{tab1}. The rest--frame
2500~\AA\ flux density we use to derive the $\alpha_{\rm{ox}}$ and
$\Delta\alpha_{\rm{ox}}$ values is extrapolated from the the nearest
available OM filter using a spectral index of $\alpha_\nu = -0.70$ (Wu
et al. 2001).

\begin{figure}
\begin{center}
\includegraphics[width=0.33\textwidth,height=0.45\textwidth,angle=-90]{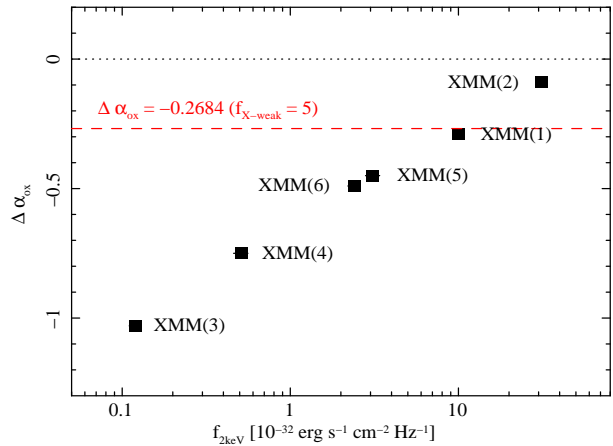}
\caption{The 2000-2010 $\Delta\alpha_{\rm{ox}}$ of PHL~1092 as a
  function of its 2~keV flux density showing visually that the
  $\alpha_{\rm{ox}}$ variation is entirely due to the long--term
  X--ray variability of PHL~1092. The level $\Delta\alpha_{\rm{ox}} =
  0$ corresponds to $\alpha_{\rm{ox}} = -1.48$, which is that expected
  for a quasar with the 2500~\AA\ monochromatic luminosity of
  PHL~1092. We also show as reference (dashed line) the level
  $\Delta\alpha_{\rm{ox}} = -0.2684$, corresponding to a quasar which
  is X--ray weak by a factor of f$_{\rm{X-weak}} = 5$. The largest
  variation is between observations XMM(2) and XMM(3) which differ by
  a factor $\sim$~260 at 2~keV (see Table~\ref{tab1}). PHL~1092 was
  X--ray weak by a factor f$_{\rm{X-weak}} \simeq 480$ during XMM(3).}
\label{histDalpha}
\end{center}
\end{figure}

As mentioned above, PHL~1092 became suddenly extremely X--ray
weak in 2008, reaching  $\alpha_{\rm{ox}} = -2.51$, to
be compared with  $\alpha_{\rm{ox}} = -1.57$ in 2003 (Miniutti et
al. 2009b). Our monitoring observations of PHL~1092 show that the
source has recovered part of its X--ray flux since 2008. PHL~1092 is still
significantly X--ray weak, but less extreme, with $\alpha_{\rm{ox}} =
-1.97$ in our latest observation performed in December 2010. In
Fig.~\ref{histDalpha} we show $\Delta\alpha_{\rm{ox}}$ as a function
of the 2~keV monochromatic flux. The almost perfect linear
correlation with $\log f_{\rm{2keV}}$ shows that the
$\alpha_{\rm{ox}}$ variability of PHL~1092 can be entirely explained
in terms of its long--term X--ray flux changes. The UV flux is
almost constant (within 10\% of its mean value), indicating
that the large $\alpha_{\rm{ox}}$ variability of PHL~1092
has little to do with mass accretion rate variations, and is instead
related to physical processes affecting the X--ray part of the SED
only. Notice that $\Delta\alpha_{\rm{ox}}$ can be used to obtain the
factor f$_{\rm{X-weak}}$ by which a given quasar is X--ray weak via
f$_{\rm{X-weak}} = 10^{-\Delta\alpha_{\rm{ox}}/0.384}$. We then infer
that PHL~1092 was X--ray weak by a factor f$_{\rm{X-weak}} \simeq 480$
in its lowest X--ray flux state as observed during XMM(3) in January 2008.

\section{UV and optical spectra}
\label{optUV}

As mentioned above, we obtained optical spectra with the Hobby--Eberly
Telescope (HET) quasi--simultaneously with observation XMM(3) and
between observations XMM(5) and XMM(6), on 2008 January 30 and 2010
October 31, respectively. The 2008 observation was also complemented
by a second optical spectrum taken on 2008 February 14 at the William
Herschel Telescope (WHT) to cover the Mg~\textsc{ii} region. Below, we
start our discussion of the optical/UV properties of PHL~1092 with a
qualitative analysis of the properties of the {\it Hubble Space
  Telescope} (HST) spectrum of PHL~1092 obtained in 2003.

\subsection{The HST UV spectrum}

\begin{figure}
\begin{center}
\includegraphics[width=0.33\textwidth,height=0.45\textwidth,angle=-90]{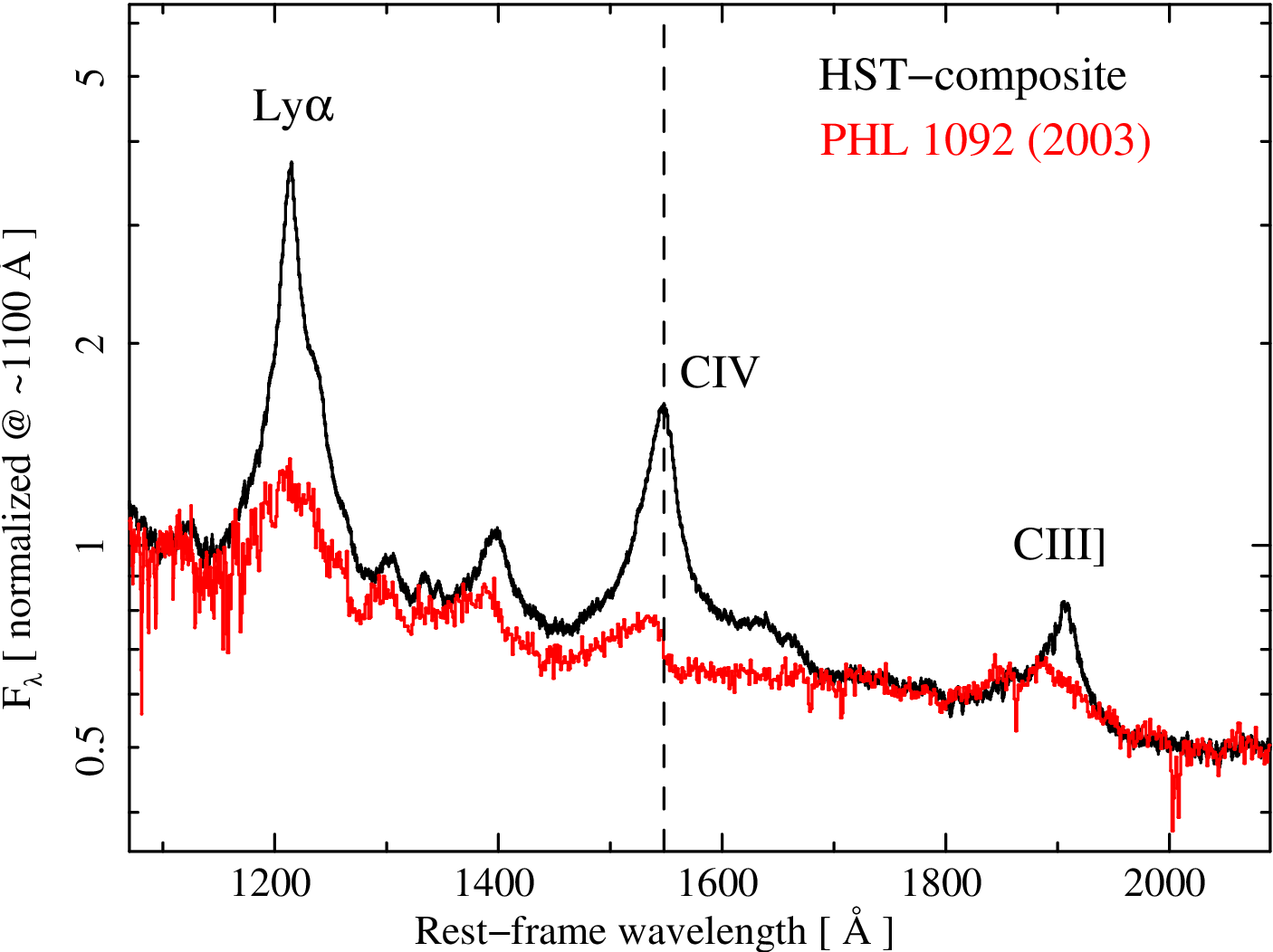}
{\vspace{0.5cm}}
\includegraphics[width=0.33\textwidth,height=0.45\textwidth,angle=-90]{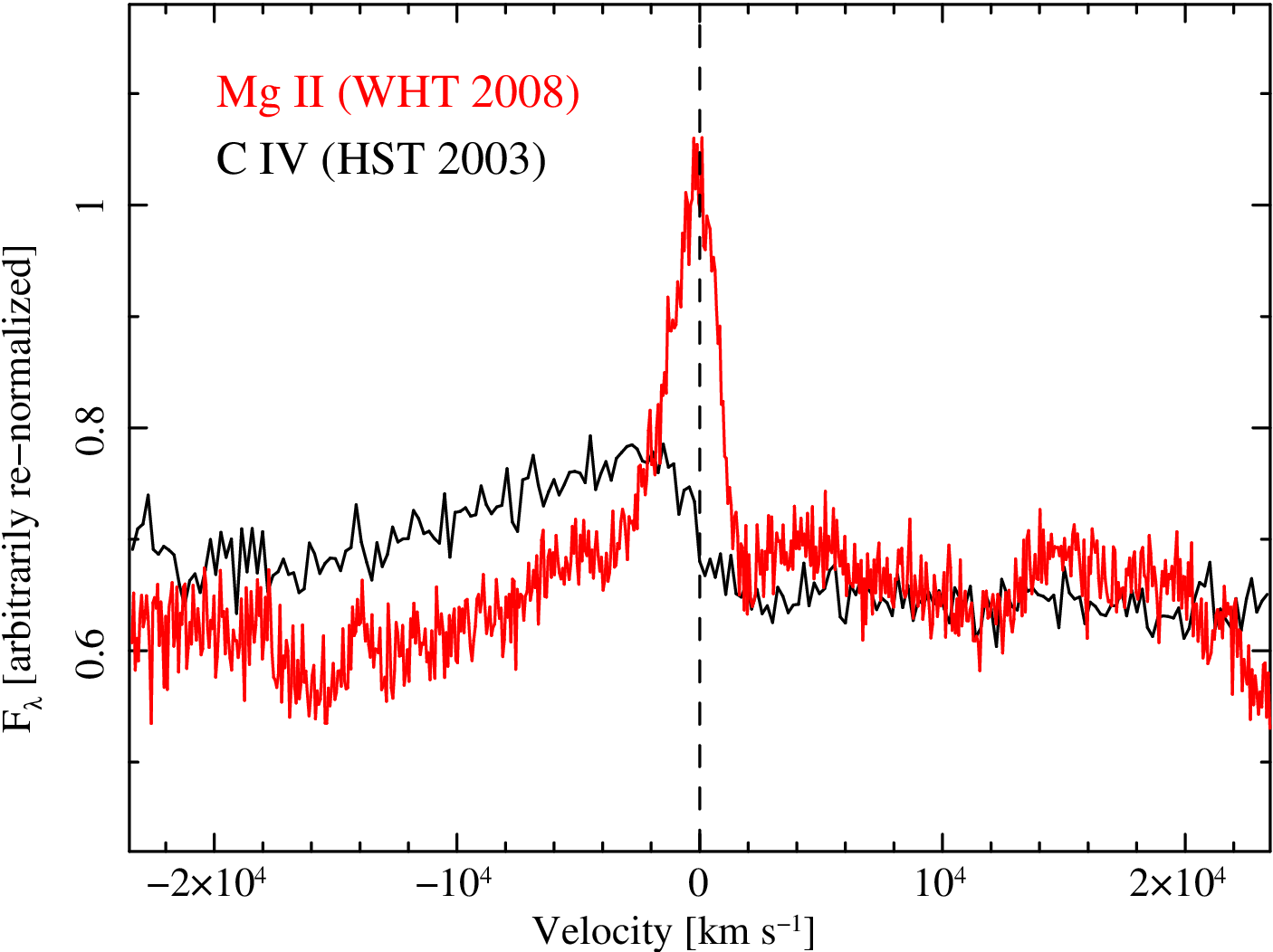}
\caption{{\bf{Top:}} The HST--STIS spectrum of PHL~1092 is compared
  with the HST--composite spectrum of radio--quiet quasars. Notice
  especially the weak, broad, and blueshifted C~\textsc{iv} emission
  line of PHL~1092. We show as reference the rest--frame C~\textsc{iv}
  wavelength of 1548~\AA\ as a dashed line. {\bf{Bottom:}} Comparison
  between the C~\textsc{iv} and Mg~\textsc{ii} line profiles obtained
  in 2003 (HST) and 2008 (WHT) respectively. The zero velocity is
  shown for reference as a dashed line.}
\label{HSTcomp}
\end{center}
\end{figure}

PHL~1092 was observed with the HST Space Telescope Imaging
Spectrograph (STIS,) on 2003 August 20 and 2003 September 18, about
one and two months after the XMM(2) observation. The observations were
performed with the G140L and G230L gratings covering the
1150--3180~\AA\ spectral range. No significant variability is detected
between the two HST spectra. Here we do not perform any detailed
scientific analysis of the HST spectrum, but use it for illustration
purposes only. In the upper panel of Fig.~\ref{HSTcomp} we show the
Ly$\alpha$ to C~\textsc{iii}] portion of the spectrum of PHL~1092 and
of the HST composite for radio--quiet sources (Telfer et
al. 2002). The high--ionisation C~\textsc{iv} emission line of
PHL~1092 is much weaker than in the average quasar. The C~\textsc{iv}
line is also broad and highly blueshifted with virtually no
contribution at the line's rest--frame wavelength (see Wu et
al. 2011). Inspection of the optical spectrum of PHL~1092 taken at the
WHT on 2008 February 14 reveals that the low--ionisation
Mg~\textsc{ii} emission line is instead centred at its rest--frame
wavelength and is relatively narrow, with a likely smaller
contribution from a broad blueshifted component. The bottom panel of
Fig.~\ref{HSTcomp} shows a comparison of the C~\textsc{iv} and
Mg~\textsc{ii} line profiles in velocity space. The two observations
are not simultaneous. C~\textsc{iv} is from the 2003 HST observation
while Mg~\textsc{ii} is from our 2008 February 14 observation with the
WHT.

The purpose of showing here the HST spectrum of PHL~1092 and the
remarkably different line profiles of C~\textsc{iv} and Mg~\textsc{ii}
is to associate PHL~1092 with other likely similar sources. The
unusual UV emission--line properties of PHL~1092 match those of the
prototypical intrinsically X--ray weak quasar PHL~1811 (Leighly et
al. 2007a, 2007b), and of the two strong Fe~\textsc{ii} emitters
and relatively X--ray weak NLS1 galaxies 1H~0707--495 and
IRAS~13224--3809 (Leighly \& Moore 2004; Leighly 2004). Leighly (2004)
associates the broad, blueshifted, high--ionisation lines
(e.g. C~\textsc{iv}) with a wind, while the narrow, rest-frame,
low--ionisation lines are likely produced in the denser disc or at the
base of the wind itself. The lack (or extreme weakness) of any narrow
C~\textsc{iv} emission line at the rest--frame wavelength in all these
sources suggests that the continuum irradiating the narrow and
low--ionisation line--emitting region has been filtered through the
wind, thus supporting the idea that Mg~\textsc{ii} and similar
low--ionisation lines are emitted further away than the wind launching
site, which Leighly estimates to be $\sim 5\times 10^3~r_g$ for
1H~0707--495 and IRAS~13224--3809.

Most of the peculiar optical/UV properties of PHL~1811,
1H~0707--495 and IRAS~13224--3809 can be explained with a
soft ionising SED depleted of X--ray photons and, at least in the case
of the two latter NLS1 galaxies, by enhanced metallicity (Leighly 2004; Leighly
et al. 2007b). The situation seems similar in
PHL~1092: we observe weak, blueshifted C~\textsc{iv} emission, likely
produced in a wind. No narrow C~\textsc{iv} emission line at
rest--frame wavelength is observed, while Mg~\textsc{ii} is relatively
strong and narrow, supporting a filtered continuum in PHL~1092 as
well. 

The C~\textsc{iv} emission line properties of PHL~1092 strongly
suggest the presence of a wind in August/September 2003, i.e. the
epoch of the HST observation. Considering the wind launching radius
estimated by Leighly for 1H~0707--495 and IRAS~13224--3809, the
physical distance of the wind from the centre is $\sim 50$~light days
in PHL~1092.\footnote{We assume here a black hole mass of $3\times
  10^8~M_\odot$, see e.g. Niko{\l}ajuk, Czerny \& Gurynowicz (2009).}
Hence, taking into account this timescale, the C~\textsc{iv} emission
corresponds to a nuclear SED which is likely similar to that we
observe in July 2003 during XMM(2), when PHL~1092 was not X--ray weak
($\Delta\alpha_{\rm{ox}}\sim -0.09$). We conclude that a wind in
PHL~1092 can likely be launched even if the source is only moderately
(or even not at all) X--ray weak.

\subsubsection{Evidence for C~\textsc{iv} variability?}
\label{Cvar}

All UV fluxes obtained with the OM during our monitoring show signs of
modest long--term variability, with a tendency of lower variability
amplitude at longer wavelengths (see Table~\ref{tab1}). The last three
{\it{XMM--Newton}} observations have exposures in multiple
filters. The variability trend above 1655~\AA\ (M2 filter) is for a
higher flux in XMM(4) a decrease in XMM(5) and a final rise in
XMM(6). On the other hand, the trend for the shortest wavelength W2
filter is, surprisingly, the opposite. It is crucial to note that the
OM filter W2 has an effective rest--frame wavelength of 1519~\AA,
which coincides with the wing of the (blueshifted) C~\textsc{iv}
emission line of PHL~1092.

\begin{figure}
\begin{center}
\includegraphics[width=0.33\textwidth,height=0.45\textwidth,angle=-90]{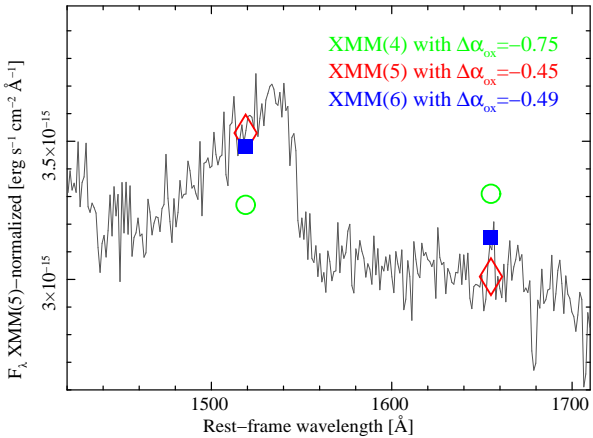}
\caption{The HST spectrum of PHL~1092 normalised to the XMM(5) OM
  photometric data around the C~\textsc{iv} emission line is shown
  with the OM monochromatic fluxes at 1519~\AA\ and 1655~\AA,
  corresponding to observations XMM(4), XMM(5), and XMM(6), which are
  the only ones to have multiple--filters exposures. The
  $\Delta\alpha_{\rm{ox}}$ for each observation is also reported as
  reference. We use open circles for XMM(4), open diamonds for XMM(5),
  and filled squares for XMM(6).}
\label{PHL_OM}
\end{center}
\end{figure}

In Fig.~\ref{PHL_OM} we show the HST spectrum of PHL~1092 in the
restricted region around the C~\textsc{iv} emission line on which we
over--plot the monochromatic OM fluxes from the W2 (1519\AA) and M2
(1655\AA) filters for the three observations with multiple--filter
exposures. As the HST spectrum was obtained at a time when PHL~1092
was most likely not particularly X--ray weak, we have re--normalised
the HST spectrum to the OM photometry of the least X--ray weak
observation XMM(5). As can be seen, the HST spectral shape is
consistent with the OM photometry during XMM(5) and XMM(6), as shown
by the open diamonds and filled squares in Fig.~\ref{PHL_OM}. On the
other hand, this is not the case for the X--ray weakest observation
XMM(4), in which the two OM filters have a flux consistent with each
other (open circles in the Figure).

This comparison suggests significant UV spectral variability. In
particular, our result suggests that the C~\textsc{iv} emission line
was either much fainter or more blueshifted during the X--ray weak
state observation XMM(4). Both possibilities (and of course a
combination of them) can explain the difference between the HST
spectral shape and the OM photometry during XMM(4). In fact, there are
numerous suggestions in the literature that the C~\textsc{iv} line
equivalent width (EW) is anti--correlated with its blueshift
(e.g. Richards et al. 2011) and that sources with smaller
C~\textsc{iv} EW (and/or larger blueshift) are typically X--ray weaker
or {\it soft--spectrum} objects (Wu et al. 2009; Kruczek et al. 2011;
Richards et al. 2011). Our result suggests that the idea that X--ray
weaker quasars have lower C~\textsc{iv} EW and larger C~\textsc{iv}
blueshift possibly holds not only statistically in large AGN samples,
but also in individual sources such as PHL~1092.  Unfortunately, the
lack of multiple--filter exposures during the X--ray weakest
observation XMM(3) prevents any further analysis of the issue. A UV
spectrum of PHL~1092 during an extreme X--ray weak epoch would clearly
solve this important issue.

\subsection{Optical HET spectra: comparison between the 2008 and 2010 observation}

The 2008 HET and WHT optical spectra of PHL~1092 from Mg~\textsc{ii}
to H$\beta$ were presented and qualitatively discussed by Miniutti et
al. (2009b). The main conclusion was that, despite the enormous X--ray
flux drop, the optical spectra did not show significant changes with
respect to older spectra (e.g. Bergeron \& Kunth 1980). Here we
compare the two HET optical spectra of PHL~1092 which were obtained in
2008 and 2010. The X--ray flux density at 2~keV increased by a factor
$\sim$~20 from 2008 to 2010.

\begin{figure}
\begin{center}
\includegraphics[width=0.33\textwidth,height=0.45\textwidth,angle=-90]{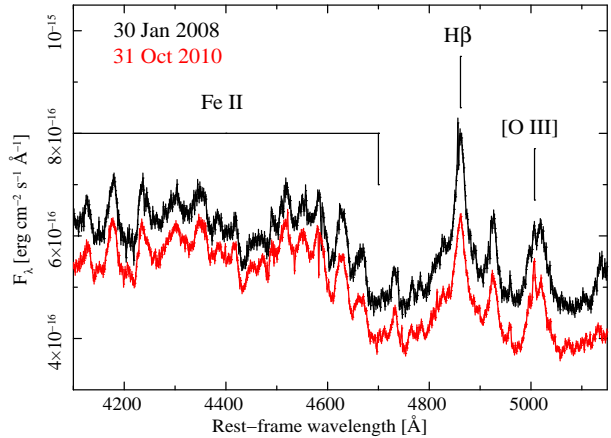}
\caption{The HET spectra of PHL~1092 from the 2008 and 2010
  observations. We do not show here the WHT data (obtained just 15
  days after the 2008 HET data) as the WHT and HET spectra are
  consistent with each other in the common spectral range.}
\label{het12}
\end{center}
\end{figure}

Spectra of PHL 1092 were taken on two occasions (2008 January 28 and
2010 October 31) with the Low Resolution Spectrograph (LRS; Hill et al
1998) on the HET (Ramsey et al 1998, Shetrone et al 2007).  On each
night two 600 s exposures were obtained with a 1.0$''$ slit and the
600 line mm$^{-1}$ grating. The calibrated spectra, which have a
resolution of $\approx$~5~\AA, are displayed in Fig.~\ref{het12} in a
restricted region covering some of the Fe~\textsc{ii} complex as well
as the H$\beta$ and [O~\textsc{iii}] emission lines. The spectra were
placed on an approximate absolute flux scale by combining photometry
taken by the Sloan Digital Sky Survey (SDSS; York et al 2000) with the
$R$-band images taken by the LRS immediately before the acquisition of
the spectra.  SDSS photometry in the $ugriz$ filters (Fukugita et al
1996) of the field containing PHL~1092 was obtained on 2004 December
14.  We derived the $R$ to $r$ transformation using field stars
(employing $r-i$ as the colour term). As shown in Fig.~\ref{het12},
the two HET spectra are quite similar in continuum shape and line
properties; the major change during this 2.8~yr span (2~yr in the
quasar rest--frame) is a lowering of the flux level by 0.14~mag
accompanied by a similar drop of the H$\beta$ line flux. This overall
flux change is roughly consistent with that observed in the UV from
the quasi--simultaneous OM photometry (see Table~\ref{tab1}).

\section{X--ray spectral analysis I: basic spectral characterisation}

\begin{figure}
\begin{center}
\includegraphics[width=0.33\textwidth,height=0.45\textwidth,angle=-90]{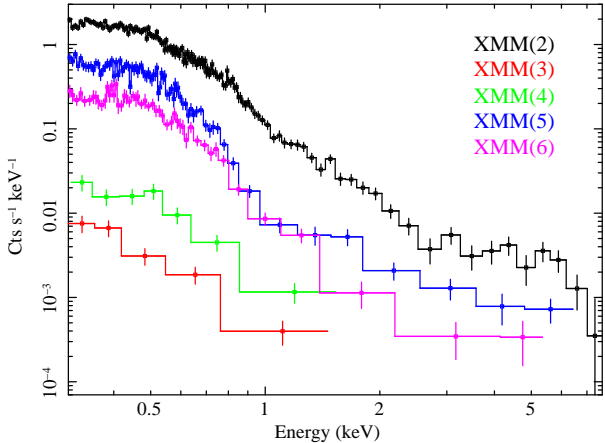}
\caption{The EPIC--pn spectra of PHL~1092 used in our analysis from
  2003 to 2010. All data are considered down to 0.3~keV and up to the
  observation--dependent energy at which the source is confidently
  detected above the background level. Observation XMM(1) is not
  considered here as no spectral data are available. The x--axis
  energy is in the observed frame here and in all subsequent figures,
  unless specified otherwise.}
\label{XMMspec}
\end{center}
\end{figure}

Having established that the extreme $\alpha_{\rm{ox}}$ variation of
PHL~1092 is due to its long--term X--ray variability only (see
Fig.~\ref{histDalpha}), we turn to the {\it XMM--Newton} data with the
goal to explain simultaneously the UV--to--X--ray SED of PHL~1092 at
all observed flux levels. As mentioned, no X--ray spectral information
is available for XMM(1), so we perform our analysis considering
observations XMM(2) to XMM(6) spanning 7.5~yr (5.4~yr in the
rest-frame) in the activity of PHL~1092 and more than two orders of
magnitude in soft X--ray flux. All EPIC--pn data are considered between
0.3~keV and up to the energy at which the source is confidently
detected above the background. The EPIC--pn spectra used in our
analysis are shown in Fig.~\ref{XMMspec}.

We first perform a basic phenomenological spectral analysis of the
X--ray data, which are all fitted simultaneously here as well as in
all subsequent Sections. As discussed by Gallo et al. (2004), Dasgupta
et al. (2004), and Miniutti et al. (2009b), the X--ray spectrum of
PHL~1092 can be described by the combination of a high--energy power
law plus a soft excess component. Here we adopt a standard
multi--colour blackbody (the {\footnotesize{DISKBB}} model, see
e.g. Mitsuda et al. 1984) and power--law model to describe the spectra
in all observations, i.e. at all X--ray flux levels. We assume
Galactic absorption with column density fixed to the value taken from
Kalberla et al. (2005), namely $3.57\times 10^{20}$~cm$^{-2}$. As the
power law slope cannot be constrained reliably in all observations we
(arbitrarily) force the photon index to be the same in all
observations. We obtain a fair phenomenological representation of the
spectra ($\chi^2/{\rm{dof}} = 425/357$) with $\Gamma = 2.0\pm
0.3$. The blackbody has a temperature in the range of $\sim
0.08-0.12$~keV, consistent with the soft excess temperature of type~1
AGN (e.g. Czerny et al. 2003; Gierli\'{n}ski \& Done 2004; Miniutti et
al. 2009a). We stress here that the blackbody component should not be
identified with thermal disc emission, as extrapolation down to the UV
underestimates the OM data by more than 2 orders of magnitude. The
results of this phenomenological description of the X--ray data are
reported in Table~\ref{pheno}. In the following Sections, we explore a
set of more physically--motivated spectral models applied
simultaneously to the EPIC and OM data.

\section{X--ray spectral analysis II: baseline model}
\label{baseline}

We adopt as our baseline model the recently developed {\footnotesize{OPTXAGN}}
model (see Done et al. 2012; Jin et al. 2012 for a detailed
description and applications). This model assumes that the disc emits,
at each radius, a colour--temperature corrected blackbody down to a
given X--ray corona outer radius R$_{\rm c}$. The main model assumption
is that the energy can no longer thermalise within R$_{\rm
  c}$, and is instead distributed between powering the soft X--ray excess
component and the high energy power--law tail via Comptonization in a
two--phase plasma. The energy distribution within the corona is
determined by the fraction of coronal luminosity powering the
high--energy power--law $f_{\rm{pl}}$ (the remaining coronal
luminosity $1-f_{\rm{pl}}$ powers the soft excess). As mentioned,
R$_{\rm c}$ sets the X--ray corona outer radius, while its inner
boundary coincides with the innermost stable circular orbit (ISCO)
$R_{\rm{isco}}$. The total coronal luminosity is proportional to
$1-R_{\rm{isco}}/R_{\rm c}$ so that no X--ray emission is produced if
$R_{\rm c}\leq R_{\rm{isco}}$.

\subsection{Soft excess as Comptonization}

\begin{figure}
\begin{center}
\includegraphics[width=0.33\textwidth,height=0.45\textwidth,angle=-90]{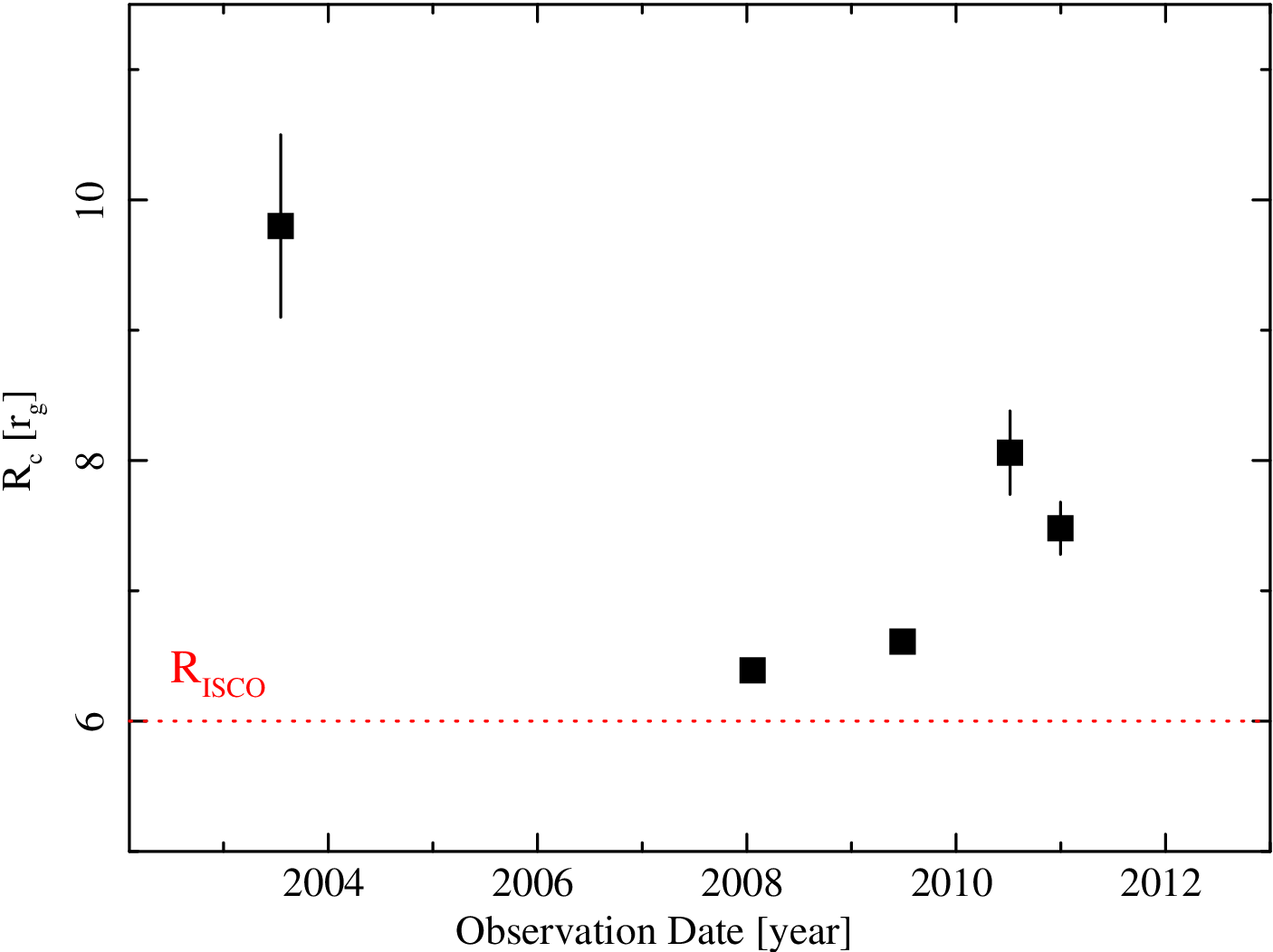}
{\vspace{0.5cm}}
\includegraphics[width=0.33\textwidth,height=0.45\textwidth,angle=-90]{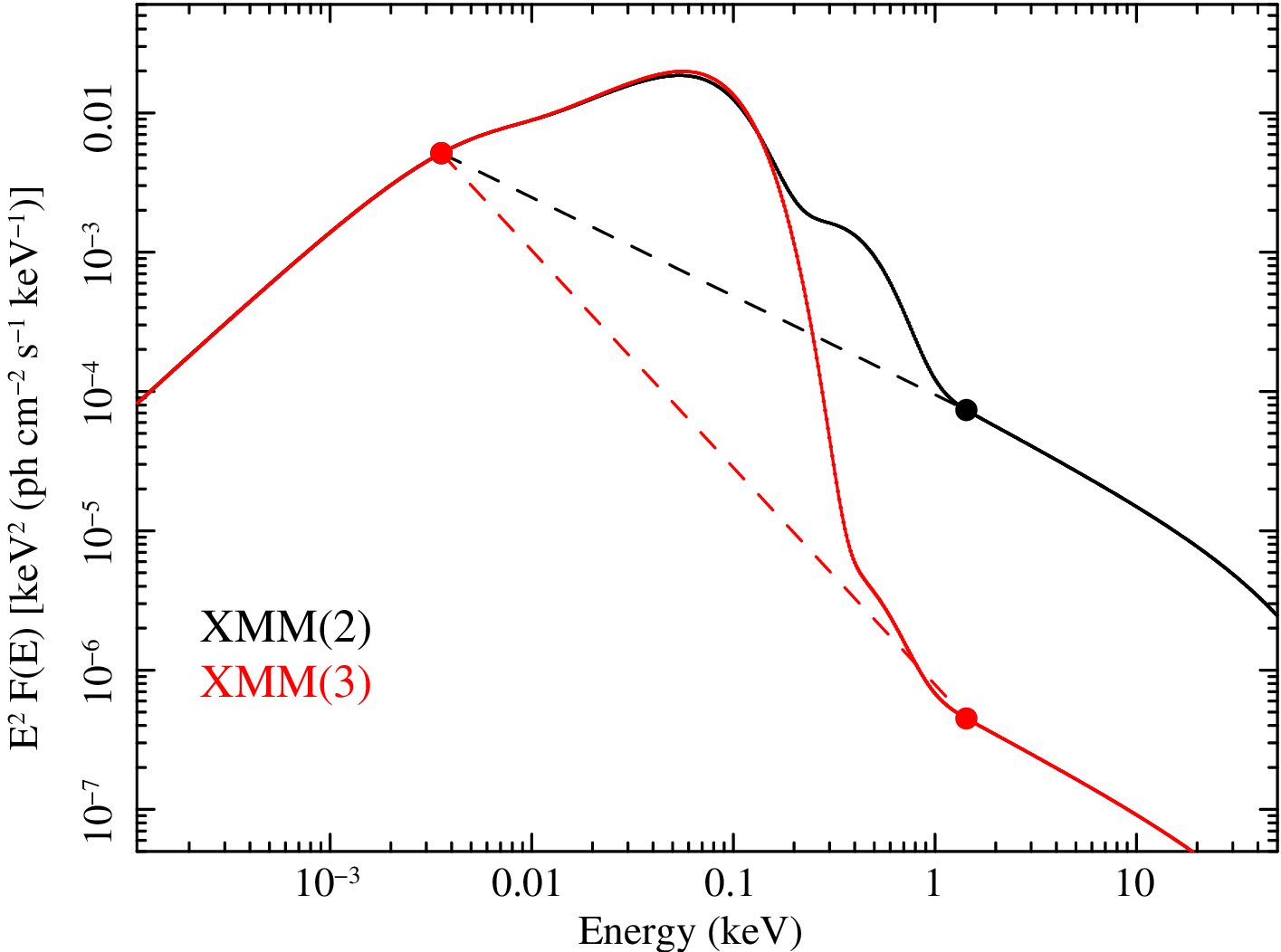}
\caption{{\bf{Top:}} The X--ray corona outer boundary R$_{\rm c}$ as a
  function of time in PHL~1092 for the case of a Schwarzschild BH with
  $a=0$ and R$_{\rm{isco}} = 6~r_g$. The collapse of the corona down
  to $\sim$~R$_{\rm{isco}}$ is able to explain the factor of $\sim$~260 X--ray
  flux drop between the highest and lowest flux observations (2003 and
  2008) at fixed Eddington ratio. {\bf{Bottom:}} The E$^{2}$F(E) model
  SED of PHL~1092 during the highest and lowest flux observations
  XMM(2) and XMM(3) as obtained with the baseline model for $a=0$ (see
  Table~\ref{mod1}, top). The SED has been corrected for Galactic
  absorption. We also show as filled circles the position of
  2500~\AA\ and 2~keV in the rest--frame. The dashed lines represent
  the optical to X--ray spectral slope for the two observations.}
\label{Figmod1}
\end{center}
\end{figure}

The first spectral model we consider is the baseline
{\footnotesize{OPTXAGN}} model in which the soft excess is due to the
Comptonization of the soft disc thermal photons in an optically thick
plasma. As the disc thermal component contributes in the UV, all the
available OM data are used simultaneously with the EPIC spectra to
constrain the thermal disc emission. As we fit simultaneously UV and
X--ray data, we include the {\footnotesize{REDDEN}} (for UV data) and
{\footnotesize{PHABS}} (for X--ray data) models to account for
line--of--sight Galactic absorption. We use the standard dust--to--gas
conversion E(B$-$V)$=1.7\times10^{-22}$N$_{\rm H}$ (Bessel 1991) to
link the UV and X--ray absorption.

Our goal is to reproduce the UV/X--ray data in all observations with
variations of a minimum number of parameters. As the UV flux of
PHL~1092 does not vary significantly (see Table~\ref{tab1}), we force
the Eddington ratio to be the same at all X--ray flux levels, i.e. in
all observations\footnote{Our assumption was checked
  a--posteriori. Indeed, if we instead assume observation--dependent
  Eddington ratios, they all appear to be consistent within the errors
  with that obtained by forcing it to be the same in all observations
  (Table~\ref{mod1}).}. Given that the phenomenological model
explored above indicates a similar spectral shape of the soft excess
in all observations (see Table~\ref{pheno}), we also assume that the
electron temperature (kT$_{\rm e}$) and optical depth ($\tau_{\rm e}$)
of the optically--thick plasma producing the soft excess are
constant. One of the model parameters is the BH spin, and we consider
two different cases, namely a non--rotating Schwarzschild BH ($a=0$)
and a maximally spinning Kerr BH ($a=0.998$). The difference between
the two cases is that the innermost stable circular orbit is
$R_{\rm{isco}} = 6~r_g$ for $a=0$ and $1.24~r_g$ for $a=0.998$ (where
$r_g =GM_{\rm{BH}} / c^2$).

We are thus left with the following free parameters: the BH mass, the
Eddington ratio, $kT_{\rm e}$, $\tau_{\rm e}$, $\Gamma_{\rm h}$,
the outer X--ray corona radius R$_{\rm c}$, and the fraction of coronal
luminosity powering the hard power--law $f_{\rm{pl}}$. However, all
these parameters are forced to be the same in all observations
except R$_{\rm c}$, $\Gamma_{\rm h}$,
and $f_{\rm{pl}}$. After a few initial tests, we found that
the data are best represented if $\Gamma_{\rm h}$ is forced to be the
same in the first three and last two observations.

We obtain a fair (although not excellent) simultaneous description of
the OM and EPIC--pn data for both the $a=0$ and $a=0.998$ cases, with
$\chi^2/{\rm{dof}}= 480/370$. The results are reported in
Table~\ref{mod1}. We obtain a BH mass of $\sim 2.4\times 10^8~M_\odot$
for the $a=0$ case and  $\sim 1.7\times 10^9~M_\odot$ for the
$a=0.998$ one. The corresponding Eddington ratios imply
super--Eddington accretion for a Schwarzschild BH with L/L$_{\rm{EDD}}
\sim 2.45$, and a sub--Eddington ratio of L/L$_{\rm{EDD}} \sim 0.31$
for a Kerr one. Using the standard mass scaling relations
(e.g. Vestergaard \& Peterson 2006) together with the optical
luminosity and the FWHM of H$\beta$ obtained from single--epoch
optical spectra, one estimates a black hole mass of $\sim 3\times
10^8~M_\odot$ for PHL~1092, in line with our best--fitting result for
the Schwarzschild BH case (Niko{\l}ajuk, Czerny \& Gurynowicz 2009).

The most striking results of our analysis are that i) the X-ray corona
of PHL~1092 is extremely compact at all probed X--ray flux levels with
$R_{\rm{c}}\leq 1.8\times R_{\rm{isco}}$ in all observations, and ii) the extreme
observed X--ray and $\alpha_{\rm{ox}}$ variability can be explained
 in terms of a variation of the X--ray corona size and hence
X--ray output (proportional to $1-R_{\rm{isco}}/R_{\rm c}$). The deep
minimum X--ray flux state observed during XMM(3) is the result of an
almost complete collapse of the X--ray corona down to the inner disc
boundary $R_{\rm{isco}}$ in both the Schwarzschild and Kerr cases.

The evolution of R$_{\rm c}$ is shown in the top panel of
Fig.~\ref{Figmod1} for the Schwarzschild BH case\footnote{The Kerr BH
  case is qualitatively the same and we discuss here the Schwarzschild
  case only for simplicity. In fact, the evolution of R$_{\rm c}$ in the
  two cases is almost identical, once expressed in units of
  $R_{\rm{isco}}$.}. In the bottom panel, we show the (unabsorbed)
model SED showing visually the difference in the corona X--ray output
at the two extreme flux levels, accounting for the large variation of
$\alpha_{\rm{ox}}$ from $-1.57$ to $-2.51$. The nominal power--law
connecting 2500~\AA\ and 2~keV in the rest--frame is also shown as a
dashed line in both cases. Since we assume that the Eddington ratio is
constant, the bolometric luminosity is the same at all X--ray flux
levels, namely $L_{\rm{Bol}} \sim 4.7 \times
10^{46}$~erg~s$^{-1}$. This implies an enormous change in the X--ray
bolometric correction ($K_X$), from $K_X \sim 8 \times 10^2$ in the
highest flux XMM(2) observation to $K_X \sim 1.2\times 10^5$ in the
lowest flux XMM(3) one for the standard 2--10~keV X--ray band.

\subsection{Can X--ray absorption make PHL~1092 a standard quasar with constant $\Delta \alpha_{\rm{ox}} \sim 0$?}

The model presented above successfully explains the $\alpha_{\rm{ox}}$
variation but at the expense of invoking a {\it breathing} X--ray
corona producing an intrinsic X--ray flux change of more than two
orders of magnitude from the highest to the lowest flux state of
PHL~1092. Here we search for a solution in which the intrinsic
X--ray flux of PHL~1092 is constant and possibly similar to that
expected for a quasar with its optical luminosity. Guided by the
observational fact that X--ray weak quasars are often associated with
intrinsic UV and X--ray absorption (at least the BAL X--ray weak
sources), we assume that the X--ray flux variability of PHL~1092
is apparent and due to X--ray absorption.

\begin{figure}
\begin{center}
\includegraphics[width=0.2\textwidth,height=0.45\textwidth,angle=-90]{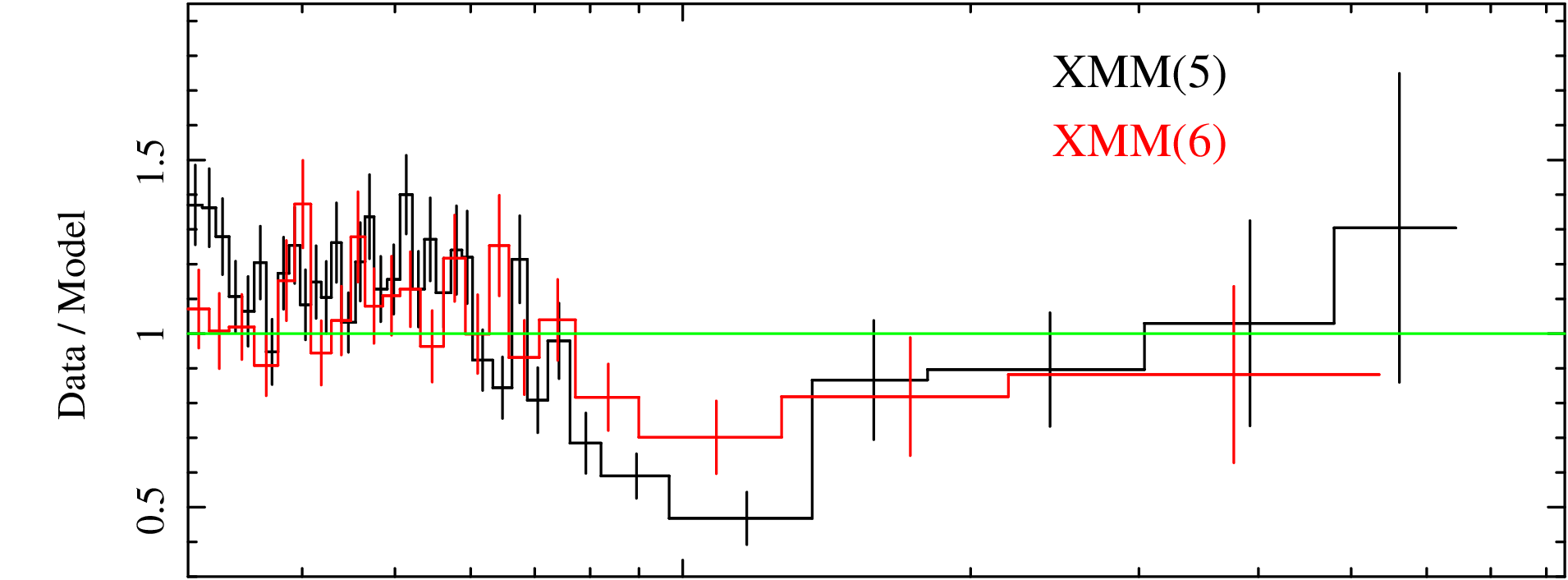}
                {\vspace{0.2cm}}
                \includegraphics[width=0.23\textwidth,height=0.45\textwidth,angle=-90]{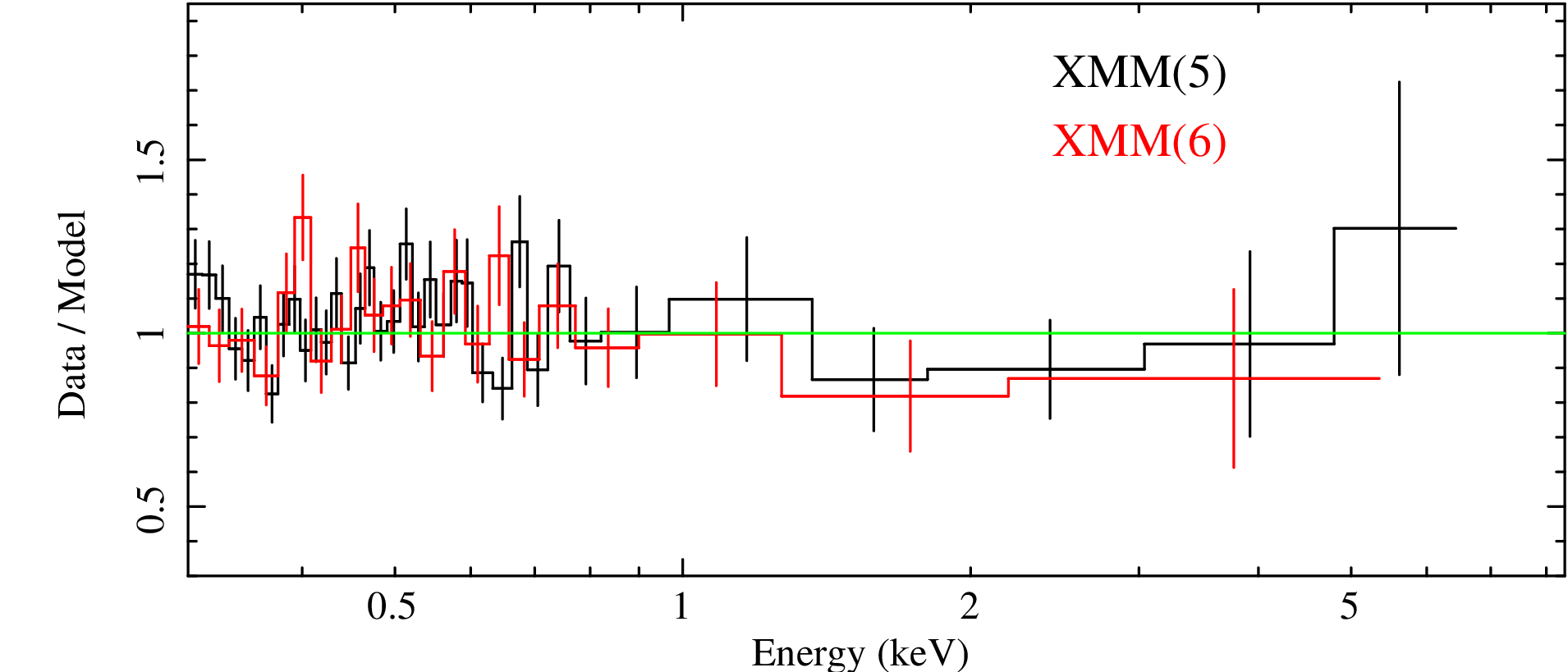}
                \caption{{\bf{Top:}} The data--to--model ratio for the
                  intermediate--flux observations XMM(5) and XMM(6)
                  using a single absorption model. {\bf{Bottom:}} Same
                  as above but including a Gaussian absorption line at
                  $E_{\rm{abs}} \sim 1.05$~keV. We only show the intermediate--flux
                  observations for visual clarity as the absorption
                  feature is more prominent in these data sets (see
                  text for details). The feature is marginally visible
                  in XMM(2) as well, while the data from XMM(3) and
                  XMM(4) are inconclusive due to the relatively poor
                  quality around 1~keV. The data have been re--binned for
                  visual clarity only.}
\label{gabsratio}
\end{center}
\end{figure}

To test this scenario in PHL~1092, we use the same model as above,
adding a partially ionised X--ray absorber with no counterpart in the
UV. As our absorption spectral model (the {\footnotesize{ZXIPCF}}
model, Reeves et al. 2008) is only defined in the X--ray band, we use
the EPIC data only, fixing the BH mass and Eddington ratio to the
values derived above, so that the UV data are automatically described
by the overall model (provided that R$_{\rm{c}} \leq 20-30~r_g$). We
discuss here for simplicity the Schwarzschild $a=0$ case only
(qualitatively similar results are obtained in the Kerr case). Our
goal is to reproduce the X--ray weakness and variability of PHL~1092
with absorption effects. Hence, we force the corona outer boundary
R$_{\rm c}$ to be the same in all observations. This implies that the
intrinsic X--ray flux of PHL~1092 is always the same. For
self--consistency, the absorber ionisation state is also forced to be
the same at all flux levels although, in principle, the ionising flux
may intercept clouds with different density at different times. The
coronal parameters affecting the soft excess ($kT_{\rm e}$ and
$\tau_{\rm e}$) and the high--energy power--law tail ($\Gamma_{\rm h}$
and $f_{\rm{pl}}$) are also forced to be constant to limit further the
number of free parameters (forcing the coronal parameters to be the same in all
observations reduces the number of free parameters from 27 to 11).

After a few tests, we found that the absorber column density and
covering fraction cannot be simultaneously constrained. We then
decided to force a marginally Compton--thick column density of N$_{\rm
  H} = 10^{24}$~cm$^{-2}$ in all observations and let the covering
fraction vary. This choice is motivated by the fact that our
absorption model neglects Compton scattering, thus becoming less
robust for much higher column densities. On the other hand, after
testing different N$_{\rm H}$ values, we find that a significantly 
lower N$_{\rm H}$ cannot reproduce the observed large X-ray flux
variability with changes in covering fraction only. Changing the
column density around the chosen value slightly affects the
best--fitting ionisation state and covering fraction but not the
overall statistical quality of the fit. Hence, the derived parameters
must be taken with caution. We do not find find a fair description of
the EPIC data ($\chi^2/{\rm{dof}} = 710/362$). It should, however, be
stressed that most of the contribution to the $\chi^2$ comes from the
intermediate--flux observations XMM(5) and XMM(6), which appear to
exhibit a further absorption structure at $\sim$~1~keV ($\sim$~1.4~keV
in the rest--frame). The absorption feature is shown in terms of
data--to--model ratio for the intermediate--flux observations XMM(5)
and XMM(6) in the top panel of Fig.~\ref{gabsratio}.

To address this feature, we added a Gaussian absorption line with the
same energy and width in all observations, but we allow the absorption
line optical depth to be observation--dependent. We obtain a good
description of the data ($\chi^2/{\rm{dof}} = 410/355$) with a
relatively broad absorption line ($\sigma_{\rm{abs}} \sim 0.20$~keV)
at $E_{\rm{abs}} \sim 1.05$~keV in the observed--frame, corresponding
to $\sim 1.47$~keV in the rest--frame. The line is more prominent in
XMM(5) and XMM(6) with optical depths of $\tau_{\rm{(5)}} = 1.00\pm
0.10$ and $\tau_{\rm{(6)}} = 0.60\pm 0.10$ (see the bottom panel of
Fig.~\ref{gabsratio}). It is also detected in the highest flux
observation XMM(2) with $\tau_{\rm{(2)}} = 0.15\pm 0.03$; it is not
detected in the lowest flux observations, where the
signal--to--noise--ratio around 1~keV is, however, limited. The broad
width and energy of the absorption line suggests an identification
with the Fe~L absorption complex around 1~keV, with possible
contribution from Ne~\textsc{ix-x} and/or Mg~\textsc{xi-xii} (see
e.g. Nicastro, Fiore \& Matt 1999). Identification of the absorption
feature with Fe~L resonant absorption at $\sim 1$~keV implies an
outflow with highly relativistic velocity $v_{\rm{out}} \sim 0.3$~c.

\begin{figure}
\begin{center}
\includegraphics[width=0.33\textwidth,height=0.45\textwidth,angle=-90]{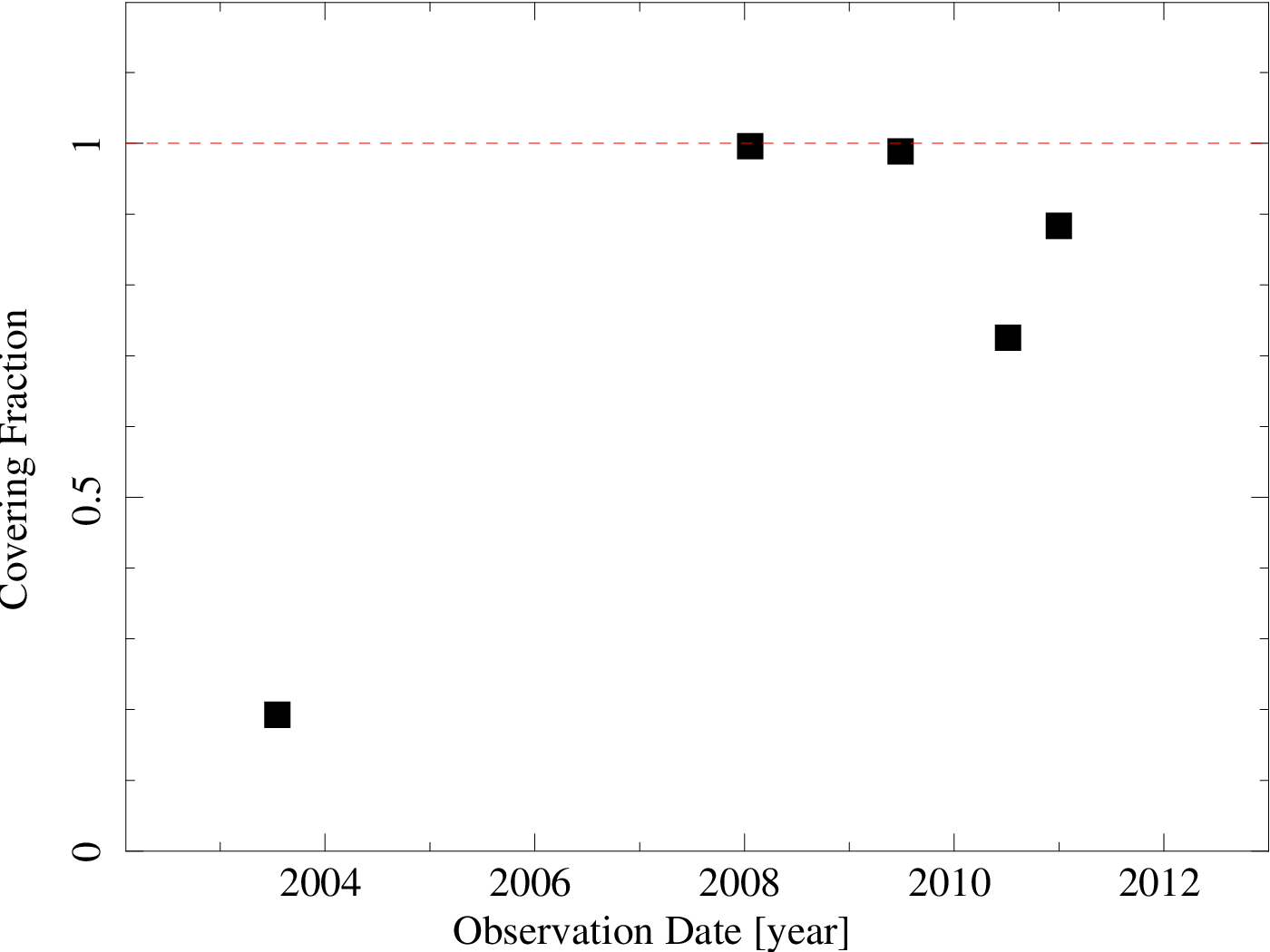}
                {\vspace{0.5cm}}
                \includegraphics[width=0.33\textwidth,height=0.45\textwidth,angle=-90]{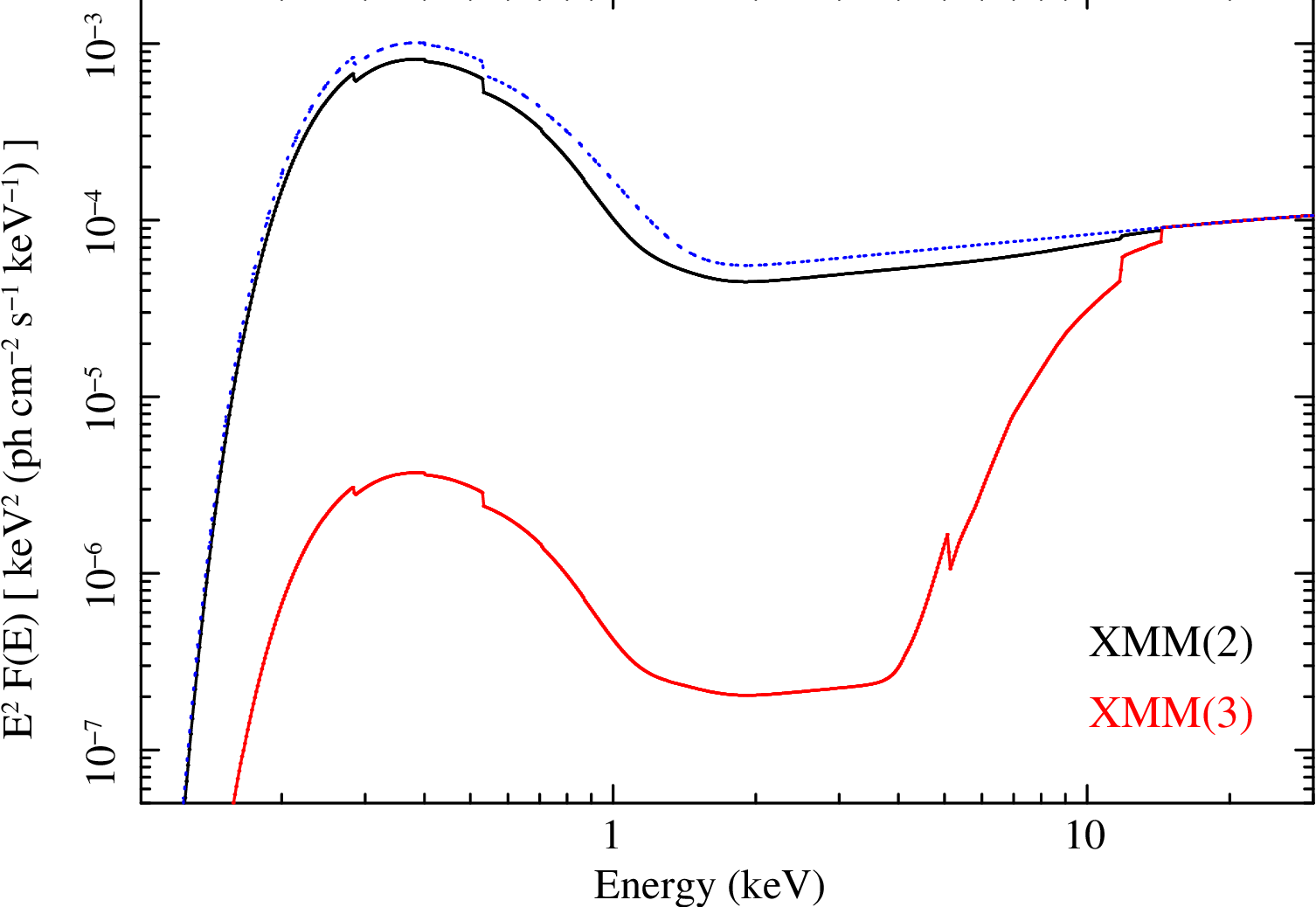}
                \caption{{\bf{Top:}} The evolution of the absorber
                  covering fraction with time. The lowest flux states
                  observed in 2008 and 2009 are due to the
                  $\sim$~100\% covering of the X--ray
                  source. {\bf{Bottom:}} The E$^{2}$F(E) model SED of
                  PHL~1092 during the highest and lowest flux
                  observations XMM(2) and XMM(3) as obtained with the
                  absorption model. The $\sim$~1~keV absorption line
                  in XMM(2) is broad and has relatively low equivalent
                  width, so it is not clearly seen on this scale. We
                  also show, with a dotted line (blue on--line), the
                  unabsorbed intrinsic X--ray flux which is the same
                  in all observations.}
\label{SEDmod2}
\end{center}
\end{figure}

On the other hand, the X-ray flux variability is produced by the
variation of the covering fraction of a relatively cold ($\log\xi \leq
-1.0$) absorber with N$_{\rm H} = 10^{24}$~cm$^{-2}$ (fixed, as
discussed above). The absorber covers $\sim$~19\% of the X--ray
source in the highest X--ray flux observation, and $\sim$~99.6\% of it
during the lowest X--ray flux one. The best--fitting parameters for
our spectral decomposition are reported in Table~\ref{mod2}. The
evolution of the cold absorber covering fraction is shown in
Fig.~\ref{SEDmod2} (top). In the bottom panel of Fig.~\ref{SEDmod2},
we show the model X--ray SED for the highest and lowest X--ray flux
states together with the intrinsic X--ray corona output (common to all
flux states).

Remarkably, if the absorber were absent, the intrinsic monochromatic
flux of PHL~1092 at 2~keV would be a factor $\sim$~1.3 higher than
that observed in the high flux state. Such unabsorbed intrinsic X--ray
flux makes PHL~1092 a standard quasar with constant $\Delta
\alpha_{\rm{ox}} = -0.05\simeq 0$. Thus, within the context of the
absorption model, PHL~1092 would be a quasar with standard X--ray
output, which only appears extremely X--ray weak at times due to
intervening absorption.

\section{X--ray spectral analysis III: the nature of the soft excess}

In the previous Sections, we have assumed that the soft X--ray excess
is produced by Compton up--scattering of the UV/EUV disc photons in an
optically thick corona with temperature $\sim 0.15$~keV. This
interpretation of the soft excess has been freqeuntly questioned by
many different authors (e.g. Gierli{\'n}ski \& Done 2004) because, in
order to reproduce the uniform spectral shape in samples of AGN with
significantly different BH mass and Eddington ratio, the electron
temperature must be remarkably uniform. The uniform soft excess shape
may indicate that processes that are independent of accretion physics
are instead at play. Below, we explore two spectral models in which
the soft X--ray excess is either an apparent feature due to
intervening absorption or a real extra emission component due to
X--ray reflection off the inner accretion disc. Both interpretations
invoke atomic processes to explain the nature of the soft excess and,
as such, they naturally provide an explanation for the uniform shape
of this feature in AGN X--ray spectra.

\subsection{Partially ionised absorption}

As mentioned above, a partial covering X--ray absorption model can
reproduce the extreme $\alpha_{\rm{ox}}$ variability of
PHL~1092. Moreover, it also explains its X--ray weakness since, if the
absorber was absent, PHL~1092 would be a quasar with standard X--ray
output and $\Delta\alpha_{\rm{ox}} \sim 0$. Here we explore the
possibility that the ionised absorber also accounts for the X--ray
soft excess. We use exactly the same model as above
but we force the X--ray corona to only power the high--energy power--law
by fixing $f_{\rm{pl}} = 1$ in all observations. This model 
eliminates the optically thick X--ray corona generating the soft
excess which has thus to be explained with a different process. In
this case we assume that the soft X--ray excess is not a real extra
emission component, but rather an apparent feature due to deep
absorption troughs affecting the intermediate X--ray energy band
($\sim 0.7-2$~keV).

The model, however, does not reproduce the data ($\chi^2/{\rm{dof}} =
750/358$). Even if the Gaussian absorption line is replaced with a
highly ionised outflowing absorber, the quality of the fit remains
unaceptable, and the power law photon index reaches an implausibly
high value of $\Gamma \sim 3.8$. In fact, the only way to obtain a fair
description of the data within this model is to allow for a large
intrinsic X--ray flux variation, i.e. for a variation in the corona
size set by R$_{\rm c}$. In this context, as the intrinsic flux
varies, we also leave the absorber(s) ionisation free to vary. This
approach produces a fit with good statistical quality
($\chi^2/{\rm{dof}} = 420/354$), and the soft excess is
reproduced. However, the model implies that the intrinsic X--ray flux
of PHL~1092 changes by a factor $\sim$~250 between the highest and
lowest flux states. Moreover, the ionisation state of the absorbers
does not follow the unabsorbed flux, although this may be understood by
considering that the ionising flux may intercept clouds of different
density and/or distance at different times. The model fails at
reproducing via absorption effects any of the observed X--ray flux and
$\alpha_{\rm{ox}}$ variability; it is just a modified version of the
baseline model. The only difference between the present and baseline 
 models is in the description of the nature of the soft
excess rather than in the mechanism behind the extreme X-ray
variability. Much higher quality existing X--ray data sets are better suited to attack the
issue of the nature of the soft excess. Hence, we do not discuss this model any
further here. We point out, however, that our absorption model does
not include re--emission, which is likely to mostly contribute in the
soft X--ray band and may be of some help in reproducing the soft
excess for specific geometries (e.g. Schurch, Done \& Proga 2009).

\subsection{Ionised disc reflection}

We now consider the possibility that the soft X--ray excess is a real
extra emission component affecting primarily the soft X--ray band. The
most attractive idea, besides Comptonization, is that ionised
reflection off the inner accretion disc contributes to the soft
X--rays. X--ray reflection off the inner accretion disc
has been shown to represent a plausible explanation of the nature of
the soft excess in many sources and, in particular, in many NLS1
galaxies (e.g. 1H~0707--495, Fabian et al. 2004, 2009; NGC~4051, Ponti et
al. 2006; IRAS~13224--3809, Ponti et al. 2010).

We apply the baseline {\footnotesize{OPTXAGN}} model as above to all
EPIC data fixing the BH mass and Eddington ratio to the best--fitting
values derived with the baseline model, depending on the adopted BH
spin. We also force the X--ray corona to only power the hard X--ray
power--law ($f_{\rm{pl}}=1$) so that the baseline model does not
produce any soft excess. To describe the soft X--ray excess as ionised
disc reflection, we use the the {\footnotesize{REFLION}} model (Ross
\& Fabian 2005) to which we apply the {\footnotesize{RDBLUR/KDBLUR}}
relativistic kernel, appropriate for a Schwarzschild/Kerr BH, and
obtained by R.~Johnstone and A.C.~Fabian from the
{\footnotesize{DISKLINE/LAOR}} models (Fabian et al. 1989; Laor 1991). As the
corona is now optically thin, the inner disc below R$_{\rm{c}}$ is
visible and we fix the inner disc radius to $R_{\rm{isco}}$ at all
flux levels.

The model free parameters are thus R$_{\rm c}$, the hard X--ray power
law photon index $\Gamma_{\rm h}$, the reflection emissivity index $q$
(where the emissivity profile is $\propto r^{-q}$), the disc
inclination $i$, and the reflection ionisation state $\xi_{\rm{ref}}$
and normalisation. However, $q$ and (obviously) $i$ are forced to be
the same at all flux levels. To limit further the number of free
parameters we attempted to force $\Gamma_{\rm h}$ to be constant, but
we found that the last three observation do require a steeper index
than the others; we thus force $\Gamma_{\rm h}$ to be the same in the
first two and last three observations, respectively. Our initial fits
show that the X--ray spectrum of PHL~1092 is completely
reflection--dominated at all flux levels. We thus force the corona
outer radius R$_{\rm c}$ to coincide with R$_{\rm isco}$ so that no
X--ray corona emission is present. 

\begin{figure}
\begin{center}
\includegraphics[width=0.33\textwidth,height=0.45\textwidth,angle=-90]{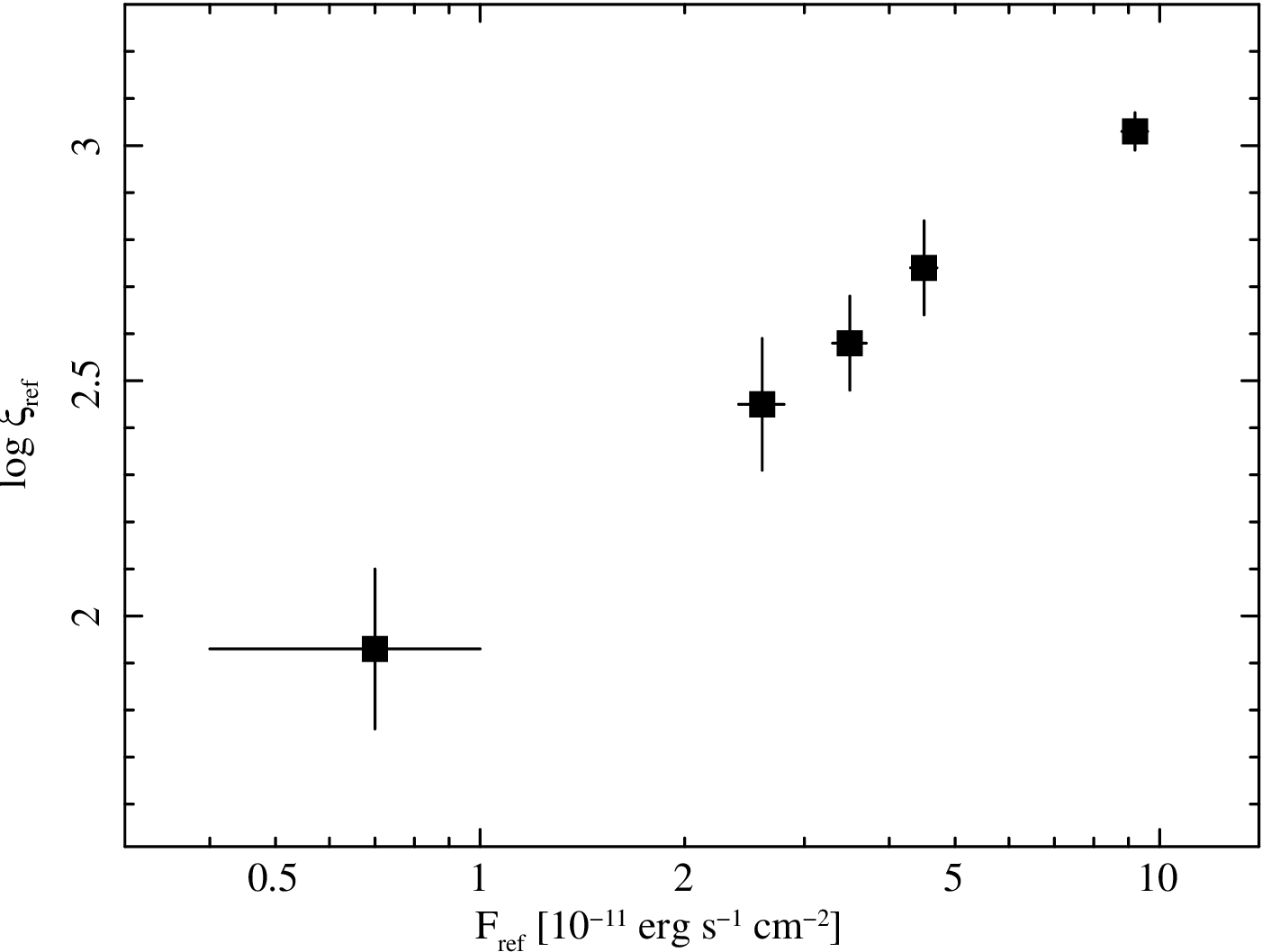}
{\vspace{0.5cm}}
\includegraphics[width=0.33\textwidth,height=0.45\textwidth,angle=-90]{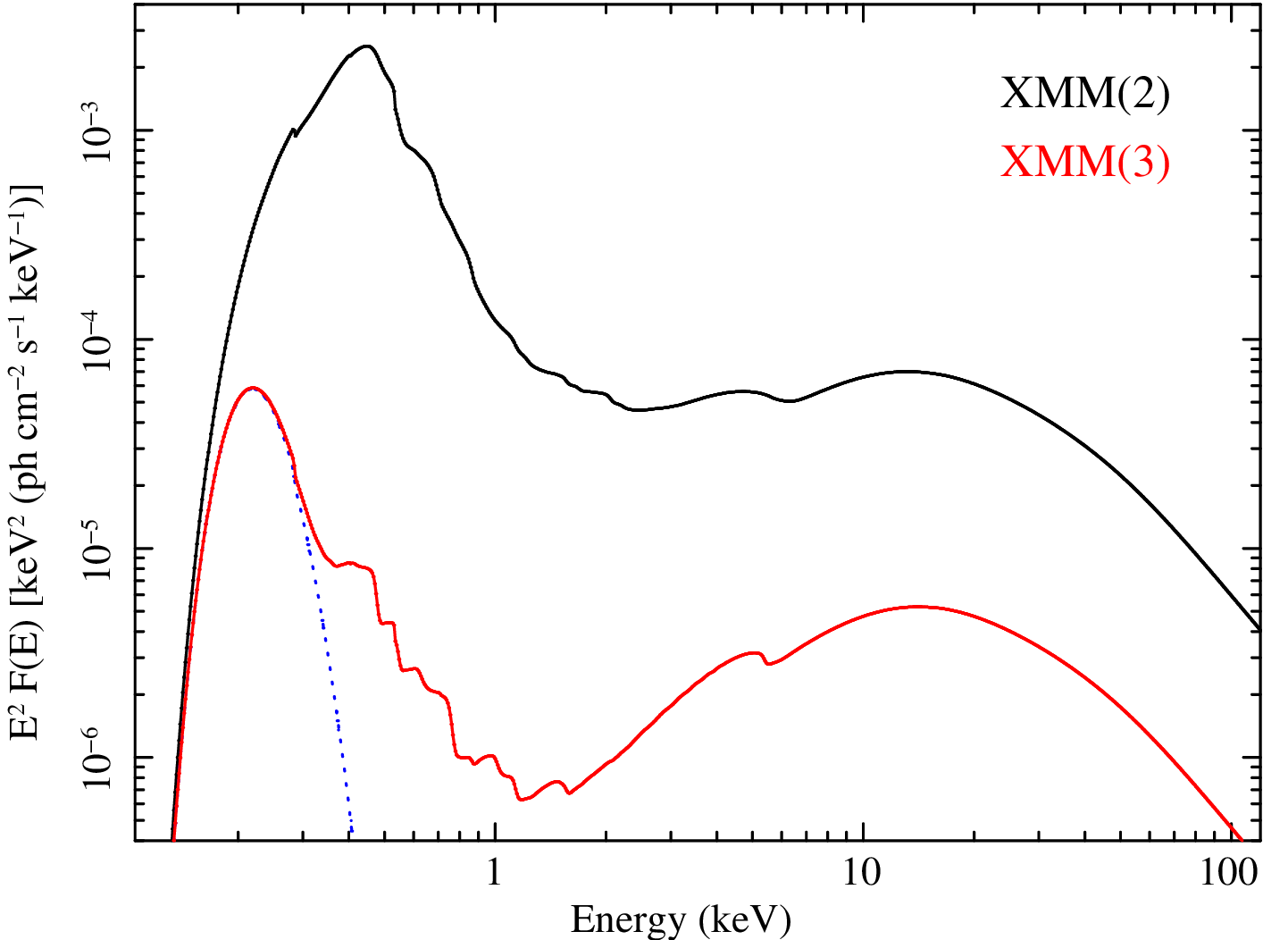}
\caption{{\bf{Top:}} The correlation between the accretion disc
  ionisation state and the total reflection flux $F_{\rm{ref}}$. Since
  $F_{\rm{ref}}$ is proportional to the flux irradiating the disc, the
  excellent correlation shows that the model self--consistently
  accounts for irradiating flux and ionisation changes. {\bf{Bottom:}}
  The E$^{2}$F(E) model X--ray SED of PHL~1092 during the highest and
  lowest flux observations XMM(2) and XMM(3) as obtained with the disc
  reflection--dominated model. The SED has been corrected for the
  intrinsic warm absorption, while Galactic absorption is present. The
  X--ray SED is completely reflection--dominated with some accretion
  disc contribution at soft X--rays (blue dashed line) which becomes
  visible in the low X--ray flux observations.}
\label{Xiref}
\end{center}
\end{figure}

We obtain a fair description of the data ($\chi^2/{\rm{dof}}=420/353$
for the Kerr BH case and $\chi^2/{\rm{dof}}=455/353$ for the
Schwarzschild BH case) with the inclusion of a
low--column/low--ionisation warm absorber. As we obtain a better
description of the data with a maximally spinning Kerr BH, we only
discuss the $a=0.998$ case here. For the warm absorber, we use the
{\footnotesize{ZXIPCF}} model fully covering the source. our results
are reported in Table~\ref{mod34}. We must point out that the warm
absorber component is not exceedingly robust as it is affected by the
presence of the thermal disc emission, which impacts the softest
X--ray energies. The latter component is not fitted but imposed by our
choice of BH mass and Eddington ratio which are, as discussed later,
somewhat uncertain. It is, however, reassuring that the warm absorber
ionisation state is well correlated with the total flux (see
Table~\ref{mod34}), ensuring the overall self--consistency of the
model. The warm absorber model improves the fitting statistics by
$\Delta\chi^2 = -32$ for 6 additional free parameters.

The column $F_{\rm{ref}}$ in Table~\ref{mod34} refers to the total
reflection flux in the 0.001--300~keV band rather than the flux in any
restricted energy band. $F_{\rm{ref}}$ is at all times proportional
(via the disc albedo $\mathcal{A}$) to the total flux irradiating the
disc $F_{\rm{irr}}$. Hence, for the model to be self--consistent,
$F_{\rm{ref}}$ must be correlated with the disc ionisation state
$\xi_{\rm{ref}}$. Fig.~\ref{Xiref} (top) demonstrates that the two
quantities are extremely well correlated, so that the reflection
scenario is indeed self--consistent. The key result of our modelling
is shown in the bottom panel of Fig.~\ref{Xiref}, where we display the
model X--ray SED for the highest and lowest flux states. The key point
is that the actual reflection flux change is only a factor $\sim$~10
between the two flux states and not the factor $\sim$~260 observed at
2~keV (see Table~\ref{tab1}). This aspect is best seen above
$\sim$~30~keV where ionisation effects are negligible. The flux change
is larger than two orders of magnitude at soft X--rays, mostly because
of the different disc ionisation. Ionisation changes affect very
significantly the reflection spectral shape, redistributing the flux
from EUV to hard X--rays. Thus, two reflectors with exactly the same
overall total flux but different ionisation may differ enormously in
the flux calculated in any restricted energy band. This is exactly
what happens here: the low flux reflector has a lower ionisation than
the high flux one (about a factor 10), which implies larger opacity in
the soft X--rays depressing the reflection flux in this band.

\subsubsection{Reflection--dominated X--ray spectrum and light--bending}
\label{lbsection}

As seen above, the X--ray spectrum of PHL~1092 can be interpreted as
being reflection--dominated at all flux levels. Reflection--dominated
spectra require the selective suppression of the intrinsic
continuum. As shown, for instance, by Martocchia, Karas \& Matt
(2000), light--bending effects can produce reflection--dominated
spectra by selectively depressing the X--ray continuum at infinity
while increasing the disc irradiation and hence the reflection
flux. According to our own ray--tracing simulations, the X--ray
spectrum becomes increasingly reflection--dominated as the X--ray
primary source location approaches the BH (see Miniutti et al. 2003;
Miniutti \& Fabian 2004). In the case of a Kerr BH, for any distance
$\leq 5~r_g$, the flux irradiating the disc is always at least twice
that escaping at infinity so that reflection--dominated spectra are
produced. We propose that the X--ray corona in PHL~1092 is confined
within a few gravitational radii from the BH and that light--bending
effects are responsible for the observed reflection--dominated
spectrum.

Let us now consider the appearance of PHL~1092 in the absence of light--bending
effects, for the standard case in which the X--ray corona is not
confined to the strong field regime close to the BH. In such a situation
50\% of the emitted X--ray flux is directed away from the disc and
detected as the X--ray continuum flux $F_{\rm{cont}}$. The remaining
50\% irradiates the disc with a flux $F_{\rm{irr}} = F_{\rm{cont}}$
and the spectrum is not reflection--dominated. $F_{\rm{irr}}$ is then
reprocessed into the X--ray reflection spectrum with flux
$F_{\rm{ref}} = \mathcal{A}~F_{\rm{irr}}$, where $\mathcal{A}$ is the
disc albedo. The remaining fraction of irradiating flux is absorbed,
thermalised, and emitted in the UV as reprocessed quasi--blackbody
emission. $\mathcal{A}$ is generally a function of the incident angle,
energy, and disc ionisation. However, for the purposes of our
discussion we assume for simplicity that $\mathcal{A} = 0.5$, i.e. in
between the typical albedo of a completely cold disc ($\sim 0.1-0.2$)
and a fully ionised one.

Light--bending effects break the equivalence $F_{\rm{irr}} =
F_{\rm{cont}}$, by differently affecting  $F_{\rm{cont}}$ and
$F_{\rm{irr}}$, but obviously maintain $F_{\rm{ref}} =
\mathcal{A}~F_{\rm{irr}}$. Let us assume for simplicity a point--like
X--ray source on the BH axis at $2~r_g$ from the BH, where
$F_{\rm{irr}}$ is amplified by a factor $\sim 1.5$ with respect to the
case with no light--bending (see e.g.  Fig.~13 of Fabian et
al. 2012). Under this assumption we can estimate the total flux that
would be observed in the absence of light--bending effects, where the
spectrum is not reflection--dominated and the disc subtends a solid
angle of $2\pi$ at the source.

It turns out that the total (continuum + reflection) flux density at
2~keV in the absence of light--bending effects is higher than the observed
(reflection only) by a factor $\sim$~70 during the low flux state and
by a factor $\sim$~9 during the high flux one. This result implies that
PHL~1092 would have had $\Delta\alpha_{\rm{ox}} \sim -0.32$ in its
lowest X--ray flux state and $\Delta\alpha_{\rm{ox}} \sim +0.28$ in
its highest. In summary, the reflection--dominated model implies that
PHL~1092 would be a quasar with relatively standard X--ray output and
$\Delta\alpha_{\rm{ox}} \sim 0$ oscillating between X--ray weak and
X--ray bright states in the absence of light--bending effects.

A final word of caution must be placed on the values of BH mass and
Eddington ratio we have assumed. We have adopted the best--fitting
values of our baseline model for a Kerr BH, but this approach is not
entirely correct. In fact, if the X--ray spectrum is
reflection--dominateed, reprocessing is inevitably important in the
optical/UV as well, as we expect that
$(1-\mathcal{A})~F_{\rm{irr}}\sim 0.5~F_{\rm{irr}}$ is absorbed and
reprocessed as thermal emission in the UV. Moreover, the X--ray
reflection spectrum itself contributes in the UV/EUV down to the
(rather uncertain) energy where Comptonization is efficient in
producing the irradiating X--ray power law. The UV outflow also
certainly contributes via scattering to the UV flux. In summary, we
caution that the BH mass estimated with our baseline model could
actually over--estimate the true value because our model attributes
the UV emission entirely to the thermal accretion disc, while other
contributions are possible (if not likely).

\section{A closer look at the  X--ray variability of PHL~1092}

\begin{figure}
\begin{center}
\includegraphics[width=0.295\textwidth,height=0.45\textwidth,angle=-90]{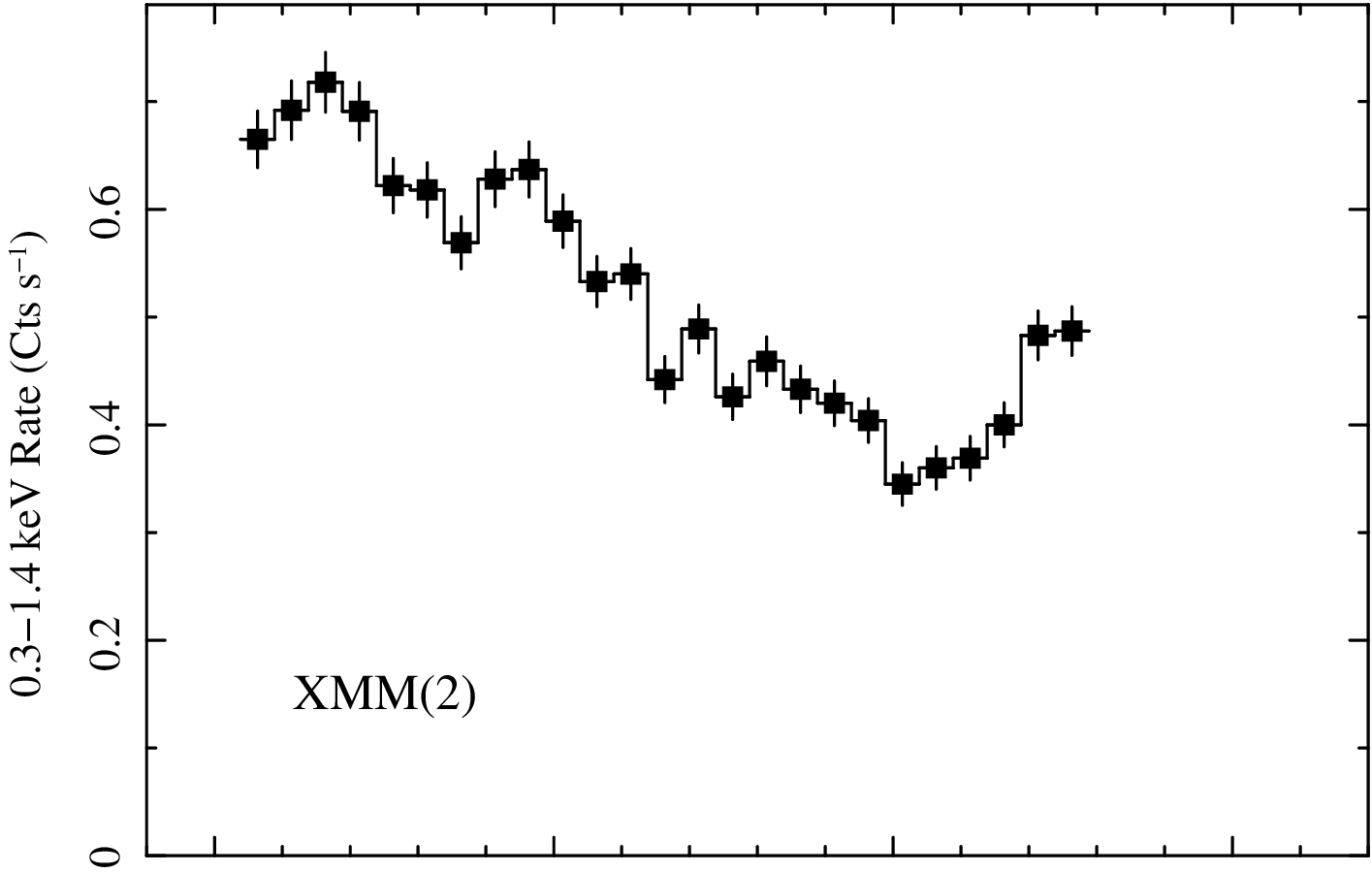}
                {\vspace{0.1cm}}
\includegraphics[width=0.295\textwidth,height=0.45\textwidth,angle=-90]{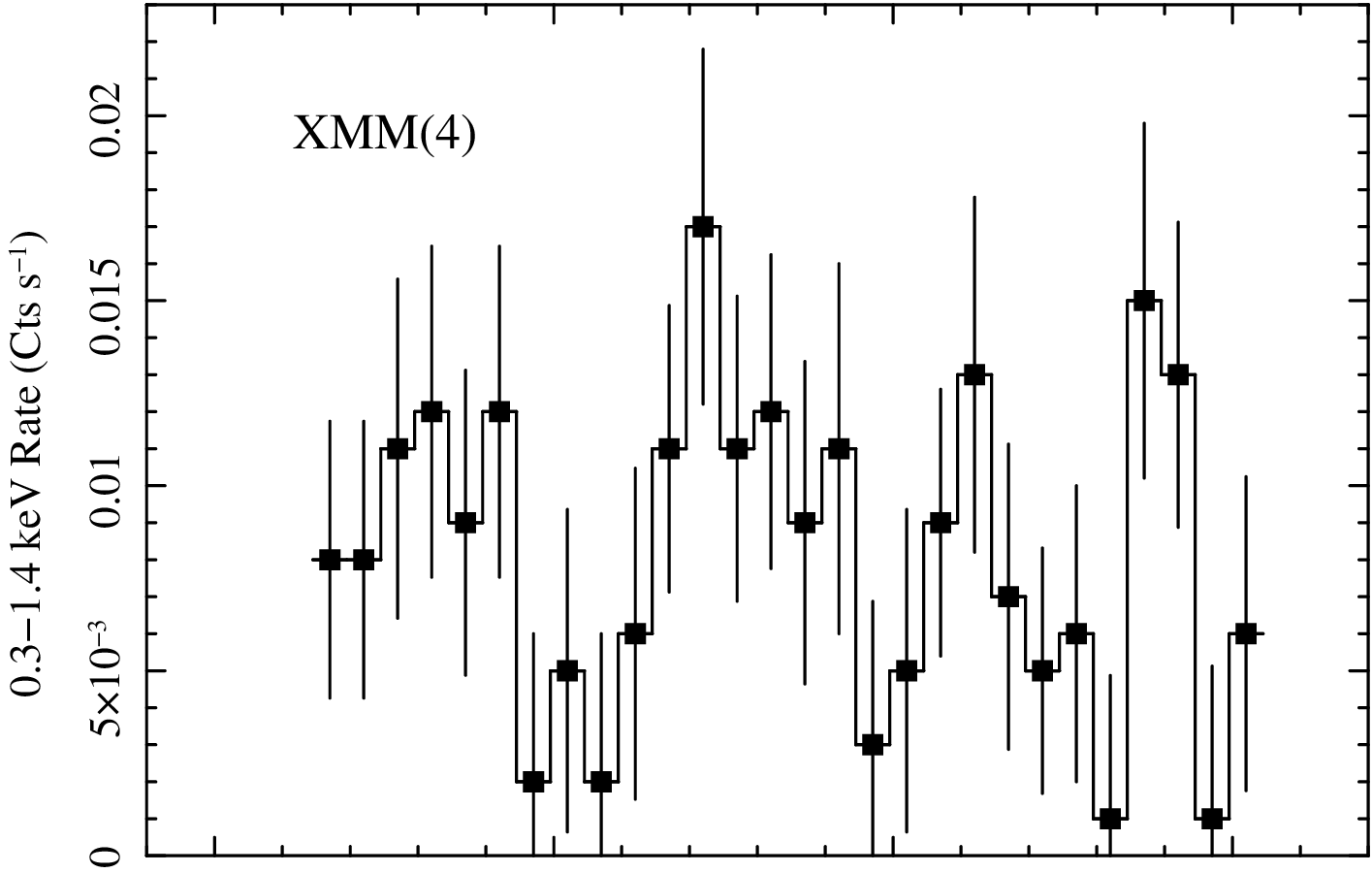}
{\vspace{0.1cm}}
\includegraphics[width=0.33\textwidth,height=0.45\textwidth,angle=-90]{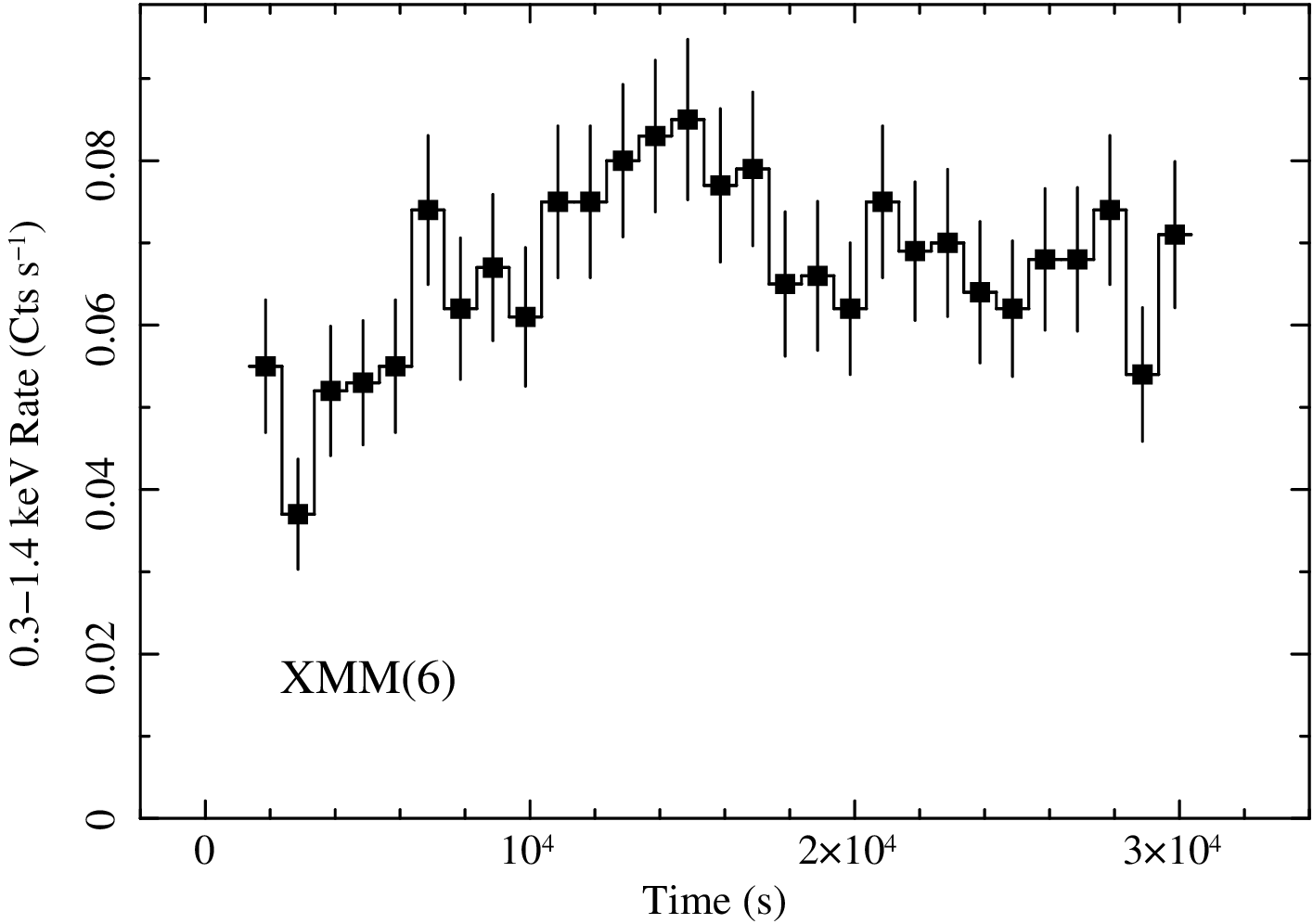}
\caption{From top to bottom we show the background--subtracted 
  light curve of PHL 1092 from the XMM(2), XMM(4), and XMM(6)
  observations. The energy band is 0.3--1.43~keV ($\sim$~0.4-2~keV in
  the rest frame) and all light curves have 1~ks bin size,
  corresponding to $\sim$~715~s in the rest frame.}
\label{xmmlcs}
\end{center}
\end{figure}

The soft X-ray light curves from three {\it XMM--Newton} observations
are shown in Fig.~\ref{xmmlcs} in the 0.3--1.43~keV band
($\sim$~0.4--2~keV in the rest--frame). The soft band is selected
because the low--flux observations are background--dominated above
$\sim$~2~keV (rest--frame) and because the count rate drops rapidly
with energy due to the steep spectrum of PHL~1092, making it difficult
to study the hard X--ray variability even at the highest X--ray flux
levels.. We show observations XMM(2), XMM(4), and XMM(6), as they are
representative of the typical X--ray variability in high, low, and
intermediate flux states, respectively. One of the key properties of
X--ray time series from accreting systems is the intimate relation
between the variability on different timescales. As discussed by many
authors (see e.g.Uttley \& McHardy 2001; Uttley, McHardy \& Vaughan
2005) the root--mean--square (RMS) variability for a given segment of
a light curve is linearly correlated with the segment mean flux on all
measured timescales. The linear RMS--flux relationship holds for the
Comptonised power--law--like component in many accreting systems, from
accreting white dwarfs to AGN, suggesting a common physical origin for
the broadband intrinsic coronal X--ray variability, independent of
source type (e.g. Uttley et al. 2004; Scaringi et al. 2012).

\begin{figure}
\begin{center}
\includegraphics[width=0.33\textwidth,height=0.45\textwidth,angle=-90]{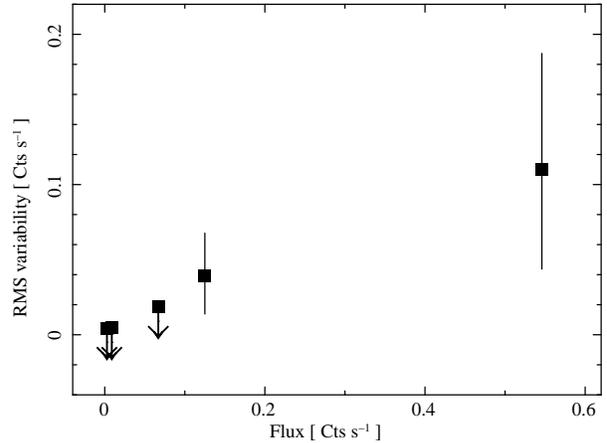}
\caption{The RMS variability as a function of flux. The RMS
  variability is computed in the frequency range $0.05-1$~mHz
  ($0.07-1.4$~mHz in the rest--frame) for all {\it XMM--Newton}
  observations.}
\label{rmsflux}
\end{center}
\end{figure}

In order to assess whether similar properties are present in PHL~1092,
we then compute the RMS variability of PHL~1092 in the frequency range
of $0.05-1$~mHz ($0.07-1.4$~mHz in the rest--frame). The 90\% error on
the RMS is computed by considering the measurement uncertainty as well
as the scatter due to the red component of the noise, using the
results in Table~1 of Vaughan et al. (2003) and following the
prescriptions of Ponti et al. (2012). The resulting RMS variability as
a function of X--ray flux (count rate) is shown in Fig.~\ref{rmsflux}
for all {\it XMM--Newton} observations. As can be seen, despite the
relatively large errors, the RMS variability appears to be correlated
with flux, in agreement with the expected linear RMS--flux
relationship (Uttley \& McHardy 2011). The size of the error bars
prevents us from claiming a robust linear RMS--flux relationship in
PHL~1092, but it is clear, for instance, that the RMS variability of
the lowest flux observation (${\mathrm{RMS}}\leq 4.2\times
10^{-3}$~Cts~s$^{-1}$ for an average count rate of $\sim 2.8\times
10^{-3}$~Cts~s$^{-1}$) is at least one order of magnitude smaller than
lower possible RMS of the highest flux one\footnote{See Table~\ref{pheno}
  for c.g.s. fluxes in the 0.5--2~keV band.} (${\mathrm{RMS}}\geq
4.4\times10^{-2}$~Cts~s$^{-1}$ for an average count rate of $\sim
0.55$~Cts~s$^{-1}$).

This result indicates that the short-- and long--timescale X--ray
variability in PHL~1092 are likely to obey the same linear
relationship as most other accreting systems. The RMS--flux
relationship we tentatively obtain in PHL~1092 may also explain the
prolonged low flux state from observations XMM(3) to XMM(4) which are
separated by $\sim$1~yr in the quasar rest--frame; the low flux state
of PHL~1092 can be seen simply as the continuation to long timescales
of the non--linear behaviour implied by the RMS--flux relationship. In
practice, as the linear RMS--flux relationship implies that the fluxes
have a log--normal distribution, AGN with large variability amplitudes
(such as PHL~1092) are more likely to exhibit excursions into
relatively prolonged low flux states than into high flux ones. This
behaviour has indeed been observed in a number of AGN and, especially,
in those with large variability amplitudes such as the NLS1 galaxies
NGC~4051 and 1H~0707--495 (see examples of prolonged low flux states
in Fig.~1 of Vaughan et al. 2011, and Fig.~1 of Fabian et al
2012). The lack of variability during prolonged low--flux states and
the extreme variability amplitude during high--flux intervals are just
opposite extremes of the linear RMS--flux relationship. As for the
frequency with which extreme flux variations can occur, Gibson \&
Brandt (2012) have shown that extreme X--ray variations ($\geq$~100\%)
such as those observed in PHL~1092 are quite rare (an upper limit of about
4\% of observations) in quasars with similar optical luminosity.

Our baseline and reflection spectral models are associated with large
and moderate intrinsic variability respectively. As such, a RMS--flux
linear relationship has to be expected. On the other hand, this is not
the case for the absorption model, as most (if not all) of the
long--term X--ray variability is attributed to absorption changes. It
remains to be seen whether a scenario in which most of the
long--timescale flux variability is induced by obscuring structures
can naturally reproduce the linear RMS--flux relation we suggest to
exist in PHL~1092.

\subsection{Long--timescale variability}

We now consider the typical rate of change of the physical parameters
governing the flux variability of PHL~1092 on long--timescales in the
framework of the three scenarios we propose.

\subsubsection{Breathing corona (baseline) scenario}

\begin{figure}
\begin{center}
\includegraphics[width=0.295\textwidth,height=0.45\textwidth,angle=-90]{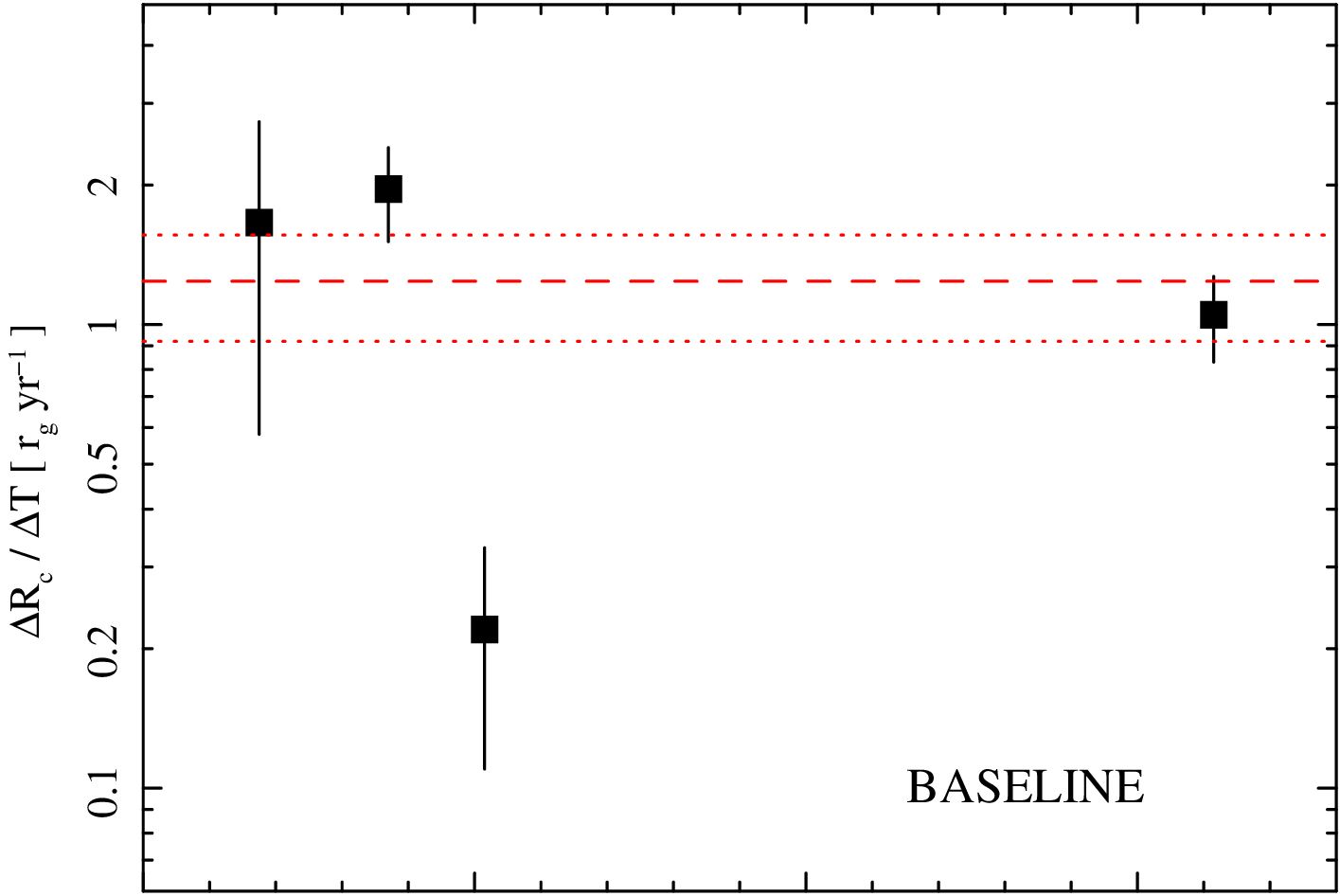}
{\vspace{0.1cm}}
\includegraphics[width=0.295\textwidth,height=0.45\textwidth,angle=-90]{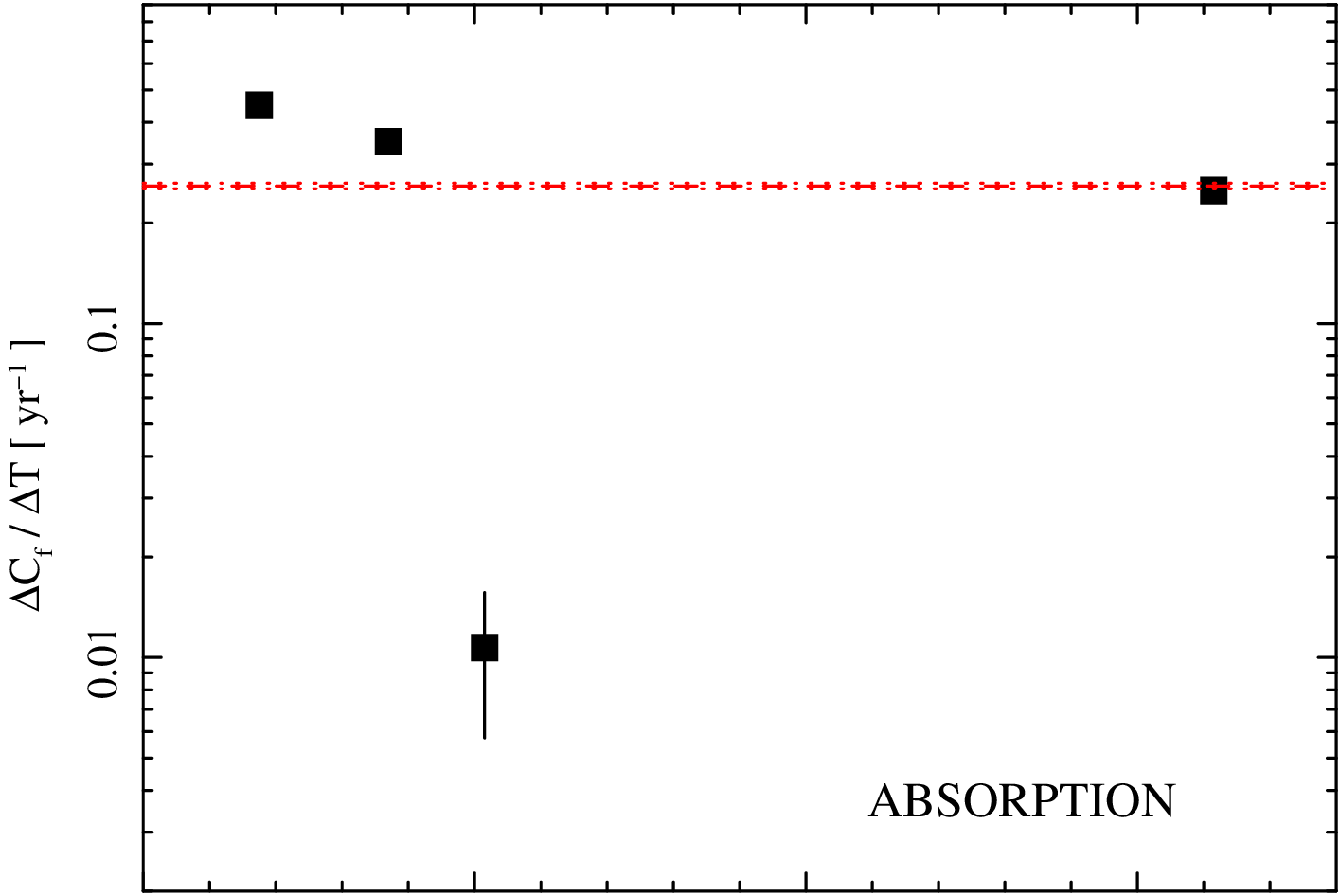}
{\vspace{0.1cm}}
\includegraphics[width=0.33\textwidth,height=0.45\textwidth,angle=-90]{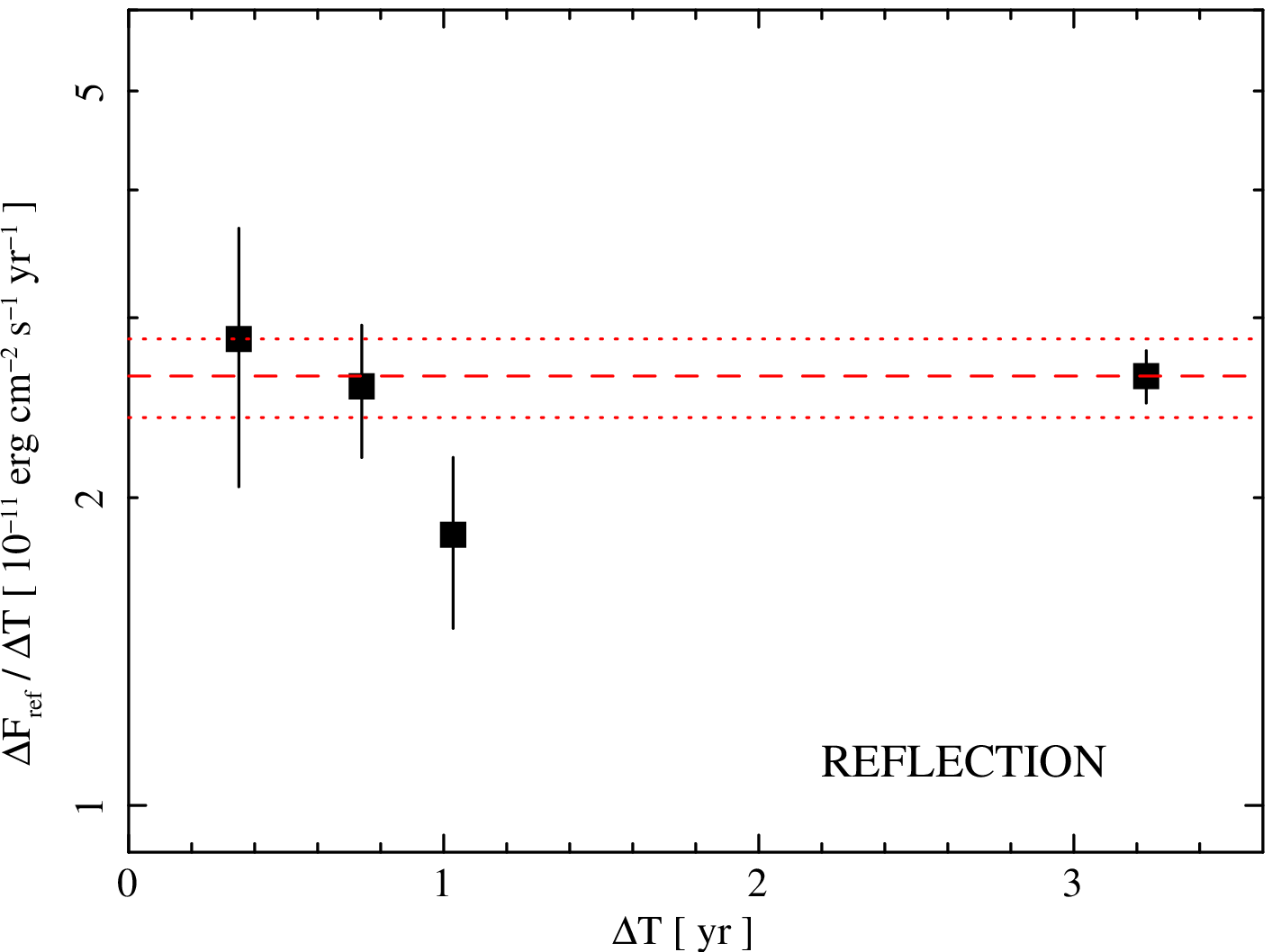}
\caption{{\bf{Top:}} The outer corona boundary rate of change
  according to the baseline {\it breathing corona} model for the
  Schwarzschild BH case. We also show the average $\langle \Delta R_c /
  \Delta T \rangle \sim 1.24~r_g$~yr$^{-1}$, which has been obtained by
  ignoring the lowest data point at $\sim 1$~yr, together with its 90\% error. $\langle \Delta R_c / \Delta T \rangle$ corresponds to a radial
  velocity of $\sim 13.7$~km~s$^{-1}$ for the adopted BH mass. {\bf
    Middle:} The covering fraction rate of change according to the
  absorption model. We also show the typical $\langle \Delta C_f / \Delta T
  \rangle \sim 0.26$~yr$^{-1}$, obtained ignoring the lowest data
  point. A better fit for the three highest data points
  can be obtained as $\Delta C_f / \Delta T \propto \Delta T^{-0.24}$
  without changing the outlier nature of the lowest data
  point. {\bf{Bottom:}} The disc--reflection total flux rate of change
  according to the disc--reflection model. $F_{\rm{ref}}$ is
  proportional to the flux irradiating the disc at all times and, as
  such, to the disc ionisation. We also show the typical flux rate of
  change $\langle \Delta F_{\rm{ref}} / \Delta T \rangle \sim 2.6\times
  10^{-11}$~erg~cm$^{-2}$~s$^{-1}$~yr$^{-1}$, obtained ignoring the
  lowest data point.}
\label{Cfrate}
\end{center}
\end{figure}

The X--ray flux is, in this model, determined by the size of the
X--ray corona. As shown in the top panel of Fig.~\ref{Figmod1}, the
baseline model suggests a {\it breathing corona} scenario in which the
corona outer boundary $R_c$ changes with time. In order to assess the
typical rate of change $\langle \Delta R_c /\Delta T \rangle$, we
define $\Delta R_c$ as the difference in corona size between two
subsequent {\it XMM--Newton} observations separated by a time span
$\Delta T$ (in the quasar rest--frame). In the top panel of
Fig.~\ref{Cfrate} we show the resulting $\Delta R_c /\Delta T$ as a
function of the rest--frame time--span $\Delta T$ for the case of a
Schwarzschild BH. The lowest data point at $\sim 1$~yr (corresponding
to the rate of change between the two lowest X--ray flux observations)
may be seen as an outlier, especially for the absorption model
discussed below. Hence, we define here the typical $\langle \Delta R_c
/ \Delta T \rangle$ as the average obtained ignoring the lowest data
point. We have $\langle \Delta R_c / \Delta T \rangle \sim
1.24~r_g$~yr$^{-1}$ which, assuming the BH mass for the Schwarzschild
case, corresponds to a radial velocity of $\sim 13.7$~km~s$^{-1}$ (we
would have $\sim 15.8$~km~s$^{-1}$ assuming a Kerr BH instead, again
showing the qualitative similarity of the two solutions). This {\it
  shrinking velocity} is likely larger than the radial drift in the
inner accretion disc (Frank, King \& Raine 1985). Hence, the corona
outer boundary is unlikely to actually physically shrink. Its apparent
motion is most likely related to a change of the active region typical
size and location rather than to real motion of the coronal outer
boundary. Notice that the lowest data point, corresponding to the
$\Delta R_c / \Delta T\sim 0.2~r_g$~yr$^{-1}$ between the two lowest
X--ray flux observations XMM(3) and XMM(4), is $\sim 6\sigma$ away
from the typical rate of change.

\subsubsection{Absorption scenario}

The covering fraction rate of change $\Delta C_f / \Delta T$ as a
function of $\Delta T$ is shown in the middle panel of
Fig.~\ref{Cfrate}. The horizontal line refers to $\langle \Delta C_f / \Delta T \rangle
  \sim 0.26$~yr$^{-1}$, obtained by excluding the lowest data point which is a clear outlier
at more than $30\sigma$. The presence of such a highly significant outlier
means that the low flux states are not only anomalous for their
extreme X--ray weakness, but also because the almost complete covering
of the X--ray source (i.e. the extreme low flux state) lasts too long
with respect to what would be expected based upon the typical covering
fraction rate of variation. In summary, the similar and high covering
fraction in two observations separated by $\sim 1$~yr has to be
regarded as coincidental, which takes away some of the appeal of the
absorption scenario. 

\subsubsection{Disc--reflection scenario}

The disc--reflection flux rate of change $\Delta F_{\rm{ref}} / \Delta
T$ as a function of $\Delta T$ is shown in the bottom panel of
Fig.~\ref{Cfrate}, where the reflection flux is computed in the
0.001--300~keV band. The typical reflection flux rate of change is
$\langle \Delta F_{\rm{ref}} / \Delta T \rangle \sim 2.7\times
10^{-11}$~erg~cm$^{-2}$~s$^{-1}$~yr$^{-1}$. The flux rate of change
between the two lowest X--ray flux observations is only $\sim 2\sigma$
away from the typical rate. Hence, all flux states, including the
extremely X--ray weak ones, are consistent with the typical
variability of PHL~1092 on long--timescales at the 2$\sigma$ level.

\subsection{Short--timescale variability}

We now consider the short--timescale ($\sim$~ks) intra--observation
variability of PHL~1092. For simplicity, we focus here on the highest
flux XMM(2) observation only, where a smooth decline of soft
X--ray flux by a factor $\sim$~2 in $\sim$~23~ks ($\sim$~16.5~ks in
the rest--frame) is seen (top panel of Fig.~\ref{xmmlcs}). This
ensures that the most stringent limits based on the short--timescale
variability can be derived since, as shown in Fig.~\ref{rmsflux}, the
RMS--flux correlation implies that the highest flux observation is
also the one with the largest RMS variability.

Let us first consider the excess variance $\sigma^2_{\rm{nxs}}$ as a
measure of the X--ray variability in XMM(2). We refer to Ponti et
al. (2012) for the operative definition of the excess variance and its
error. Using time bins of 250~s and dividing the XMM(2) light curve in
intervals of 10~ks (both in the quasar rest frame), we have
$\sigma^2_{\rm{nxs},10ks} = 0.012^{+0.063}_{-0.005}$ in the 0.4--2~keV
band (rest--frame). As shown by Ponti et al. (2012), the X--ray excess
variance in the soft and hard X--ray band are tightly correlated. If
this is the case for PHL~1092 as well, we can then make use of the
results in Table~3 of Ponti et al. (2012), where the authors report
the $\sigma^2_{\rm{nxs}}$--$M_{\rm{BH},X}$ relationship for the hard
X-ray $\sigma^2_{\rm{nxs}}$, to give an estimate of the BH mass of
PHL~1092 as derived from its X--ray variability. We obtain
$M_{\rm{BH},X} = 6^{+10}_{-5}\times 10^6~M_\odot$, in good agreement
with previous studies (Niko{\l}ajuk, Czerny \& Gurynowicz 2009). We
point out, however, that this result is obtained assuming that the
soft X-ray excess variance is indeed tightly correlated with the hard
X-ray one, as the case for the AGN in the work by Ponti et
al. (2012). However, the soft X--ray variability in PHL~1092 may not
be intrinsic (but driven, for example, by absorption changes) and may
thus over--estimate the intrinsic X--ray variability of the source. We
should then take a conservative approach and consider the
$M_{\rm{BH},X}$ value derived above as a lower limit.

As already discussed by Niko{\l}ajuk et al. (2009), the BH mass
derived from PHL~1092 X--ray--variability is about two orders of
magnitude lower than that estimated through single--epoch optical
spectra, which is $M_{\rm{BH},opt} \sim 3\times 10^8~M_\odot$. This
means that i) the short--timescale variability of PHL~1092 is
exceptionally high with respect to the AGN population or ii) the
formulae used to derive the BH mass from the optical luminosity and
the H$\beta$ FWHM do not apply as standard in the case of PHL~1092. As
a note of caution, it should be mentioned that PHL~1092 is a
wind--dominated object in the UV (see Section~\ref{optUV}). The
standard BH mass scaling relations (e.g. Vestergaard \& Peterson 2006)
are based on reverberation--mapped objects which are not
wind--dominated. Moreover, the H$\beta$ emission region likely sees a
continuum which has been filtered through the wind, which makes the
mass scaling relation possibly more uncertain in this case. Hence, the
BH mass scaling relations may not be highly reliable for
wind--dominated systems such as PHL~1092, as discussed by Richards et
al. (2011).

The BH mass derived from our UV--to--X--ray spectral analysis with the
baseline model is much higher than that derived from X--ray
variability and, in the Kerr BH case, it is also much higher than
$M_{\rm{BH},opt}$ (we derive $\sim 2.4 \times 10^8~M_\odot$ and $\sim
1.7 \times 10^9~M_\odot$ for the Schwarzschild and Kerr BH cases
respectively, see Table~\ref{mod1}). The BH mass we derive from
spectral fitting is based on the assumption that the optical/UV
emission detected with the OM is entirely due to intrinsic accretion
disc thermal emission. It is impossible, in fact, to reach the OM
fluxes with a lower BH mass and with a reasonable value for the
Eddington ratio (say lower than 5), so that the data do not seem to
allow for much lower BH masses than those reported in Table 3. The
only way to lower the BH mass from spectral fitting would be to
consider that the optical/UV fluxes are contaminated or even dominated
by a different emission component. Such component may be reprocessed
emission, likely a combination of disc thermal reprocessing and
reflection, plus scattering from the UV wind. This may be relevant for
our reflection--dominated model (which prefers the Kerr solution,
hence a likely too large BH mass) for which a significant contribution
of reprocessing to the UV flux has to be expected. In summary, if
reprocessing is important, the BH mass derived from our spectral fits
to the UV and X--ray data is likely over--estimated. However, the
lack of a self-consistent disc reprocessing model (thermal plus
reflection) from optical/UV to X--rays, and of a reliable model for the wind
scattering contribution to the UV continuum, does not allows us to
robustly infer reliable BH mass estimates within this scenario.

The short--timescale X--ray variability of PHL~1092 can also be used
to constrain the X--ray corona size under the assumption that the
variability is intrinsic (if not, the coronal size derived below has
to be considered as a lower limit). As can be seen in
Fig.~\ref{xmmlcs}, significant variability is detected down to the bin
size. In fact, considering shorter bins, we detect significant
variability down to $\Delta t = 500$~s during the XMM(2) high--flux
observation. Simple light--travel time arguments imply an X--ray
corona size of the order of $D_c = c\Delta t /(1+z)\sim 1.1 \times
10^{13}$~cm. The gravitational radius scales linearly with the BH mass
and is $1.48~M_8\times 10^{13}$~cm, where $M_8$ is the BH mass in
units of $10^8~M_\odot$, so that $D_c \simeq 0.7~r_g~M_8^{-1}$. Hence,
any BH mass exceeding $1.4\times 10^8~M_\odot$ would imply a very
compact corona with $D_c \leq 1~r_g$. Even considering the unceratinty
in BH mass discussed above, this result suggests that X--rays come
from one single compact active region at any given time, supporting
the compact corona required by our reflection--dominated model.

Let us now consider the implications of the short--timescale
variability of PHL~1092 for the three proposed scenarios. We assume
here that the same mechanism responsible for the long--timescale
variability acts also on short--timescales. This is of course a strong
assumption, and we cannot exclude that two different processes are at
play on different timescales. It is however instructive to explore the
consequences of such an assumption on the proposed
spectral/variability models.

\subsubsection{Breathing corona (baseline) scenario}

In the {\it breathing corona} scenario, the observed XMM(2)
short--timescale variability can be produced by a smooth contraction
of the corona outer boundary from $R_{\rm{c}}\sim 10.3~r_g$ to $\sim
8.9~r_g$ (we discuss the Schwarzschild case only for simplicity). As
this happens in 16.5~ks in the rest--frame, the short timescale
variability implies an unphysically large radial velocity of $\sim
0.3$~c. This highly relativistic radial velocity of the corona outer
boundary reinforces our conclusion that the X--ray corona cannot
physically shrink. The measured $R_c$ is much more likely to represent
the typical size (and/or distance from the BH) of the X--ray active
region at any given time. For example, the rate of change of $R_c$
could represent the velocity with which the magnetic
structures responsible for heating the coronal plasma can re--organise
themselves, leading to the activation of X--ray emitting regions
of different size and/or location.

\subsubsection{Absorption scenario}

For the absorption model, the observed variability can be obtained
with a covering fraction change of $\Delta C_f \sim$~50\% in $\Delta
T\sim 16.5$~ks (rest--frame). Considering an X--ray source size $D_s$
and covering from a single obscuring structure (cloud), this
corresponds to a cloud velocity of $v \sim \Delta C_f D_s / \Delta T =
(D_s / r_g) \times 1.06\times 10^{9}$~cm~s$^{-1}$ for our
best--fitting BH mass in the Schwarzschild case. Assuming a typical
X--ray source size of $10~r_g$, we infer a cloud velocity $v \sim
0.35$~c in the plane of the sky. This velocity is of the order of the
outflow velocity that we infer from our spectral analysis by identifying
the $\sim$~1 ~keV absorption feature with Fe~L absorption. Such an
outflow velocity is also in the range of that of the ultra--fast
outflows that have been claimed at Fe~K energies in a significant
fraction of radio--quiet AGN (e.g. Tombesi et al. 2010 and 2012). In
summary, if absorption is invoked to explain not only the
long--timescale but also the short--timescale variability, a
ultra--fast outflow is necessary. Such a high velocity, however, makes
it even more difficult to explain the prolonged low flux states of
observations XMM(3) and XMM(4) as clouds moving relativistically
should cover/uncover the X--ray source on very short--timescales.

\subsubsection{Disc--reflection scenario}
\label{shortref}

In the reflection--dominated scenario, the X--ray flux variability is
driven by the variability of the flux irradiating the accretion disc
together with the associated disc ionisation change. Considering our
best--fitting model for observation XMM(2), we find that the observed
soft X--ray flux variability is reproduced with a slightly smaller
variation of the irradiating flux and ionisation state, namely a
factor $\sim$~1.5 instead than a factor 2 in 16.5~ks
(rest--frame). This corresponds to $\Delta F_{\rm{ref}} / \Delta T
\sim 2\times 10^{-15}$~erg~cm$^{-2}$~s$^{-2}$. The reduction of the
intrinsic X--ray variability amplitude because of ionisation effects
also reduces the intrinsic excess variance by $\sim 50\%$, and
slightly increase the X--ray variability BH mass estimate which,
according to this model, would be of the order of $M_{\rm{BH},X} \sim
10^{7}~M_\odot$ (although with large errors).

\section{Summary of the data analysis results}
\label{discussion}

We have applied to our {\it XMM--Newton} monitoring observations of
PHL~1092 a series of models that can explain the extreme X--ray
($\alpha_{\rm{ox}}$) variability of PHL~1092 within different
contexts. Below, we summarise the results of our analysis in the
framework of the three proposed scenarios.

\subsection{Baseline model}

Our baseline model implies an intrinsic X--ray flux variation of more
than two orders of magnitude in 3.2~yr (rest--frame). Within the
framework of that model, this flux variation is associated with the
collapse of an already compact X--ray corona down to the ISCO (or
close to it). This result is obviously highly model--dependent as by
construction the only way in which the model can produce variable
X--ray output at fixed Eddington ratio is by varying the X--ray corona
size. Any model able to account for large X--ray variability
and almost constant optical/UV emission would be
successful. However, the model is reasonably self--consistent and
provides an elegant solution in terms of a breathing X--ray corona for
$\alpha_{\rm{ox}}$ variation that is entirely driven by X--ray flux
variability, as is the case in PHL~1092 (see
Fig.~\ref{histDalpha}). The model accounts for the soft excess in
terms of Comptonization of the soft disc photons in the optically
thick corona. Its electron temperature is similar to that inferred
from the analysis of the soft excess in the vast majority of type~1
AGN. As such, it remains to be seen whether this remarkably uniform
electron temperature in accreting BH spanning orders of magnitude in
BH mass and Eddington ratio can be physically explained.

In both the Schwarzschild and Kerr cases, the
observed $\Delta R_{\rm c}$ implies a radial shrinking of the X--ray
corona outer radius with a radial velocity of $\sim
10-15$~km~s$^{-1}$, increasing to $\sim 0.3$~c if the model also needs
to explain the short--timescale X--ray variability of PHL~1092. This
suggests that the corona is not actually physically shrinking, but
that the heating mechanism (likely associated with magnetic fields) is
occurring in more localised sites during low--flux states. Notice that
the very compact X--ray corona in low--flux states would also imply
that light--bending effects can no longer be neglected, likely
producing a strong reflection contribution off the inner accretion
disc. As such, the model is not entirely self--consistent. However, the
relatively limited quality of the data does not allow us to include the
disc reflection component into the baseline model without
over--modelling the X--ray data.

\subsection{Absorption model}

The absorption model is based on the idea that the observed dramatic
X--ray flux variability is not intrinsic but is due to
intervening absorption. Indeed, we find an attractive solution in
which the intrinsic X--ray flux of PHL~1092 is constant as is the UV
flux (and hence the accretion rate). The observed variability can be
entirely explained in terms of a relatively cold absorber (with
$\log\xi \leq -1.0$) with varying covering fraction. The absorber
covers $\sim$~19\% of the X--ray corona in high X--ray flux states and
almost 100\% of it in low flux ones. It should be pointed out that
such a low--ionization absorber should imprint absorption features in
the UV, which are not seen in the HST spectrum. This issue will be
discussed in Section~\ref{wider}. Remarkably, the intrinsic (constant)
X--ray flux of PHL~1092 in the absorption scenario is consistent with
that expected for a quasar with the PHL~1092 optical
luminosity. Hence, within the absorption--dominated scenario, PHL~1092
is an intrinsically standard quasar with $\Delta\alpha_{\rm{ox}} \sim
0$. An absorption feature at $\sim$~1.5~keV in the rest--frame is also
seen and cannot be produced by the relatively low--ionisation
absorber. If the feature is identified with the Fe~L complex, it
implies the presence of a second absorbing phase with relatively
high--ionisation and highly relativistic outflow velocity (of the
order of $\sim 0.3$~c).

Such a highly relativistic outflow velocity may be shared by the cold
absorber as well. In fact, if we make the strong requirement that the
model must be able to reproduce also the short-timescale X--ray variability
of PHL~1092 with covering fraction changes, a similar outflow velocity
must be invoked for the cold absorber as well. On the other hand,
such velocity makes it difficult to account naturally for the fact
that the two subsequent observations XMM(3) and XMM(4), separated by
$\sim 1$~yr in the rest--frame, share an almost identical covering fraction
(of $\sim 100$~\%). This prolonged low flux state is at odds with the
typical absorber variability properties which would predict a $\Delta
C_f \sim 0.2-0.3$ over a $\sim 1$~yr timescale.

In summary, the absorption model can explain the variability of
PHL~1092 on long--timescales but requires the extremely X--ray weak
states to be somewhat exceptional as they are clear outliers and do
not match the typical absorption variability properties on
long--timescales. The bottom line is that the low flux states last too
long to match the typical covering fraction rate of change. In fact,
excursions into relatively prolonged low flux states are a common
characteristic of AGN with large variability amplitude and obeying a
linear RMS--flux relationship, which we suggest exists in PHL~1092
as well. It remains to be seen whether a self--consistent absorption
model is able to induce a linear RMS--flux relationship, thus
explaining the prolonged low flux states in the absorption
scenario. On the other hand, the baseline and reflection models, which
are both associated with intrinsic X--ray flux variability, are
naturally coupled to the linear RMS--flux relationship (which predicts
prolonged low--flux states), so that they may be regarded as less
problematic in this respect.

\subsection{Reflection model}

A reflection--dominated model represents a plausible alternative to
the scenarios discussed above. We are able to reproduce the observed
factor $\sim$~260 variation at 2~keV by only invoking a factor
$\sim$~10 change in the intrinsic X--ray flux, coupled with associated
disc ionisation changes. The excellent correlation between the disc
ionisation and the total reflection flux (proportional to that
irradiating the disc) ensures that the model is
self--consistent. Ionised disc reflection also provides a natural
explanation for the soft X--ray excess with no need for Comptonization
in an optically thick corona, which is not the case for any of the above 
models. The model implies that PHL~1092 appears predominantly X--ray
weak because its X--ray corona is extremely compact and centrally
concentrated, within a few $r_g$ from the BH. Such compact corona also
self-consistently predicts reflection--dominated spectra because of
light--bending effects. In absence of light--bending, PHL~1092 would in
fact appear as a non--X-ray weak quasar with a residual
$\alpha_{\rm{ox}}$ variation due to to moderate intrinsic X--ray
variability (a factor 10 in 3.2~yr in the rest--frame).

The reflection model can explain both the long-- and short--timescale
X--ray variability of PHL~1092 with relatively modest intrinsic flux
changes. As the model is associated with intrinsic variability, the
(tentative) linear RMS--flux relationship we suggest to exist in
PHL~1092 comes as no surprise. Hence the relatively prolonged
low--flux state (lasting $\sim 1$~yr) can be seen simply as a
consequence of the RMS--flux linear relationship (i.e. of the
log--normal distribution of fluxes).

\section{Discussion: PHL~1092 in the wider context}
\label{wider}

PHL~1092 shares many common properties in the optical/UV with the
prototypical non--BAL X--ray weak quasar PHL~1811 (see also the case
of LBQS~0102-2713, Boller, Schady \& Heftrich 2011). Their UV spectra
are characterised by weak, broad, and asymmetric high--ionisation
emission lines and strong, narrow, low--ionisation lines at the
rest--frame wavelength. Most of their unusual optical/UV properties
can be understood by assuming a two--component
broad--emission--line--region (BELR) comprising a wind
(e.g. C~\textsc{iv}) and a disc (e.g. Mg~\textsc{ii}) component
irradiated by a soft SED depleted of X--ray photons (Leighly et
al. 2007b; Wu et al. 2011). Filtering of the ionising flux through the
wind also explains the lack of high--ionisation lines in the disc
component which is then located at a greater distance from the BH than
the wind launching site. The implication is that the BELR sees a soft
X--ray weak SED which may be intrinsic to these sources (as in our
baseline model), due to absorption, or induced by light--bending
effects.

Our baseline model associates X--ray weak sources with a compact
corona confined within a few $r_g$ from the BH. This result arises
because the X--ray output is proportional to the size of the X--ray
corona, i.e. $L_X \propto (1-R_c/R_{\rm{isco}})$. In this respect, the
baseline model resembles the disc--reflection one, as the X--ray
emission from an extremely compact corona will inevitably be
affected by light--bending effects, which further reduce the X--ray
output at infinity and imply the presence of reflection--dominated
spectra, especially if a rapidly rotating Kerr BH is considered. We
are therefore left with two main physical scenarios: PHL~1092 (and
likely other non--BAL X--ray weak sources) is characterised by i) an
unusually compact X--ray corona or ii) X--ray--only absorption.

Wu et al. (2011) recently collected a sample of 10 PHL~1811
analogs at high redshift with a mean $\Delta\alpha_{\rm{ox}}\sim
-0.40$. The X--ray analysis of this sample suggests that these objects
have an average photon index $\Gamma = 1.2\pm 0.5$, which appears to be
harder than that of standard X--ray detected AGN, although the
relatively large error bar does not allow firm constraints on the
spectral shape. If the typical spectrum of the PHL~1811 analogs is
indeed harder than usual, it may be related either to intervening
absorption or to the presence of a reflection--dominated spectrum, two
of the models that we consider here for PHL~1092 and that have been
suggested by Wu et al. (2011) for the PHL~1811 high--redshift analogs.

If absorption is invoked, the material must lie closer to the central
BH than the BELR (say that for Mg~\textsc{ii}) so that the ionising
continuum is effectively X--ray weak thus explaining the observed
optical/UV emission line properties (see Leighly 2004). On the other
hand, UV disc photons should not be affected, as no absorption is seen
in the UV spectrum. The X--ray absorber could be naturally associated
with the putative shield which is often invoked in outflow models not
to over--ionise the gas at the base of the outflow, thus enabling
acceleration (e.g. Murray et al. 1995; Wu et al. 2011). As such, the
absorber may be co--spatial with the outflow: wind simulations show
that the outflow is likely self--shielding as the column density close
to the disc surface can be high enough to enable the acceleration of a
line driven wind (e.g. Proga, Stone \& Kallman 2000; Proga \& Kallman
2004). This shielding X--ray absorber could well have been detected in
a few cases of BAL quasars (e.g. Chartas et al. 2002). If
low--ionisation high--column--density gas exists at almost all radii
close to the disc surface (as per e.g. Proga \& Kallman 2004, see
their Fig.~1), the UV emission from the extended disc would always be
affected, independently of the specific line--of--sight. Hence, such
simulations may not be totally relevant for the case of AGN with no
absorption signatures in the UV.

The lack of UV absorption in PHL~1092 (and the presence of the broad,
blueshifted C~\textsc{iv} emission line) most likely indicates that
our line--of--sight only crosses the highly--ionised outflow polar
zones, so that no absorption features are imprinted in the
UV. According to the work of Schurch, Done \& Proga (2009), which is
based on the wind simulations by Proga \& Kallman (2004), polar
lines--of--sight are characterised by high ionisation, while
equatorial ones by cold absorption (i.e. equatorial lines--of-sight
intercept the gas responsible for the shielding). A pole--on
line--of--sight into the wind funnel would indeed be consistent with
the detected blueshifted C~\textsc{iv} emission line.

Let us now consider whether the X--ray--only absorber we infer from
our absorption--model is consistent with this picture. The flux
variability and transient X--ray weakness of PHL~1092 is due to the
variation of the covering fraction of a relatively cold absorber with
$\log\xi \leq -1.0$. If the absorber also covers the UV--emitting
disc, such a low--ionisation absorber would imprint features in the UV spectrum
as well, as seen in BAL quasars but not in PHL~1092 and similar
sources. Indeed, according to Proga \& Kallman (2004), such
low--ionisation is only achieved for equatorial lines--of--sight (say
$\geq 70^\circ$, see Fig.~1 of their work). This means that we would
be intercepting the denser part of the outflow (basically the
shielding gas), which conflicts with the fact that no absorption
features are imprinted in the UV (dominated instead by broad and
blueshifted emission lines).

The presence of a cold X--ray absorber associated with the lack of UV
absorption represents a problem for the absorption scenario, as
detailed by Wu et al. (2011). The only possible way out we can
envisage is that the X--ray absorber is more compact that the
UV--emitting disc so that the UV emission is not significantly absorbed. Let
us assume an absorber confined within a fiducial radius of $\sim
50~r_g$, so that the UV emission from within that region is small and
the UV spectrum does not present any absorption feature. Our
absorption model is based on the {\footnotesize{XSTAR}} code, and the
ionisation parameter is defined as $\xi = L_{\rm{ion}}/(R^2n)$, where
$n$ is the gas number density, $R$ its distance from the ionising source, and
$L_{\rm{ion}}$ the ionising luminosity between 1 and 1000~Ry.  The
intrinsic ionising luminosity of PHL~1092 is $L_{\rm{ion}} \sim
3.5\times 10^{46}$~erg~s$^{-1}$, so that $n(R/r_g)^2=2.9\times
10^{19} \xi^{-1}$~cm$^{-1}$ (adopting the best--fitting BH mass in the
Schwarzschild case). For the ionisation parameter we derive ($\log\xi
\leq -1.0$), this implies that a compact absorber with $R\sim 50~r_g$
can only be obtained with $n \geq 10^{17}$~cm$^{-3}$, which seems
highly unlikely.

Hence the absorption model we propose is not self--consistent as the
low--ionisation gas detected in absorption (and needed to
reproduce the X--ray flux variability) cannot survive in the compact
region that is necessary to avoid significantly affecting the UV.  This
conclusion is, however, somewhat model--dependent because our
absorption model is likely over--simplified. For instance, one can
imagine a situation in which the absorbing cloud(s) have a
multi--layer structure. The outer layers are highly ionised and reduce
via absorption and Compton scattering, the ionising luminosity
irradiating the inner layers. If so, the cloud(s) core could possibly
survive with relatively low ionisation even very close to the ionising
source. This hypothesis would require consideration of a multi--layer
absorption model in which Compton scattering is self--consistently
taken into account, which also means considering with care all
possible outflow geometries/velocities. Given the limited quality of our
X-ray data, we must defer this study to future higher quality
observations of PHL~1092 or of other similar (X--ray brighter)
sources.

If a reflection--dominated spectrum is invoked, the
overall optical/UV to X--ray SED is intrinsically steep as the
ionising continuum is X--ray depleted and, as such, the unusual
optical/UV emission line properties of PHL~1092 and other similar
sources can be explained. We attribute this effect to light--bending
close to the BH which depresses very significantly the continuum flux
that can escape to an observer at infinity. Our analysis of PHL~1092
shows that a reflection--dominated spectrum due to the presence of a
very compact and centrally concentrated X--ray corona (few $r_g$ in
size) can indeed produce extreme X--ray weak states as well as
dramatic X--ray variability in the soft band by invoking much lower
amplitude variation in the intrinsic X--ray flux. Most of the soft
X--ray variability is accounted for by moderate disc ionisation
variation, associated with the similar amplitude intrinsic flux.

As mentioned above, the UV properties of PHL~1092 are similar to those
of two other X--ray weak NLS1 galaxies: 1H~0707--495 and
IRAS~13224--3809. Remarkably, the X--ray spectra of 1H~0707--495 and
IRAS~13224--3809 have been interpreted as being
disc--reflection--dominated (Boller et al. 1993, 2002, 2003; Fabian et
al. 2002, 2004, 2009, 2012; Ponti et al. 2010). In both cases, a
significantly enhanced Fe abundance is required to model the X-ray
spectrum which is in agreement with the enhanced metallicity inferred
by Leighly (2004b) from the UV spectra. A recent observation of
1H~0707--495 (Fabian et al. 2012) suggests that the X--ray corona is
confined within $\sim 2~r_g$ from the central BH during extremely low
X--ray flux states, while it is likely more extended during higher
X--ray flux states, in agreement with light--bending model predictions
(Miniutti \& Fabian 2004). Moreover, soft X--ray time lags of the
order of the light--crossing time of a few $r_g$ have been detected in
1H~0707--495, and their properties (energy and frequency dependence)
support the overall interpretation in terms of disc reflection (Fabian
et al. 2009; Zoghbi et al. 2010; Zoghbi, Uttley \& Fabian
al. 2011). It should be mentioned that, although no complete
spectral--timing model has been presented, the properties of
1H~0707--495 have been suggested to be consistent with an
absorption--dominated interpretation, which, however, may require a
special line--of --sight (Miller et al. 2010). The detection of
time--lags in several other sources appears to exclude the possibility
that we are dealing with such special lines--of--sight
(e.g. Emmanoulopoulos, McHardy \& Papadakis 2011; Tripathi et
al. 2011; Zoghbi \& Fabian 2011; Zoghbi et al. 2012). Moreover, the
time--lag amplitude appears to scale with BH mass as predicted within
the reverberation scenario if the reflector is few $r_g$ only from the
X--ray source (De Marco et al. 2012). Gibson, Brandt \& Schneider
(2008) showed that non--BAL X--ray weak quasars make up less than
about 1--2\% of the optically selected quasar population. On the other
hand, Wu et al. (2011) reported that the fraction of 
PHL~1811 analogs in the radio--quiet population is $\leq$~1.2\%. It is
reassuring that the fraction of AGN that can be very strongly
reflection--dominated is constrained to be $\leq$~2\% by the hard
X--ray cosmic background (Gandhi et al. 2007), although the cosmic
X--ray background sources are likely in a different luminosity regime
from the SDSS quasars considered by e.g. Wu et al. (2011).

\section{Conclusions}

We have analysed {\it XMM--Newton} observations of PHL~1092 covering
nearly 10~yr (7.5~yr in the rest--frame) of its activity and spanning
more than two orders of magnitude in soft X--ray flux. We complement
our analysis with optical spectra obtained in January 2008 and October
2010 and with a qualitative discussion of the UV HST spectrum from a
August/September 2003 STIS observation. The main results of our
analysis can be summarised as follows:

\begin{enumerate}

\item The remarkable $\alpha_{\rm{ox}}$ variability of PHL~1092 is
  entirely driven by the X--ray variability. During our monitoring
  campaign, we detect a maximum X--ray flux variability by a factor
  $\sim 260$ in $\sim$~3.5~yr (rest--frame), while the UV variability
  is confined to within 10--15\%. At its minimum X--ray flux level
  (January 2008), PHL~1092 is about a factor of $\sim 480$ X--ray weak
  with respect to a standard quasar with its optical luminosity.

\item The UV spectrum of PHL~1092 is characterised by weak, broad, and
  blueshifted C~\textsc{iv} emission, most likely originating in a
  wind. The lack of absorption features in the UV spectrum of PHL~1092
  most likely indicates that the outflow does not cross our
  line--of--sight, in contrast with the case of BAL quasars. It
  presents many analogies with the prototypical non--BAL X--ray weak
  quasar PHL~1811 and with the two NLS~1 galaxies 1H~0707--495 and
  IRAS~13224--3809. We tentatively report UV variability in the
  C~\textsc{iv} region based on a comparison between the single--epoch
  HST spectrum and the UV photometry from the {\it XMM--Newton} OM
  observations. The possible C~\textsc{iv} variability appears to be
  associated with X--ray weakness. Our results are consistent with the
  idea that the C~\textsc{iv} emission line is weaker and/or more
  blueshifted in X--ray weaker states. This implies either that the
  ionising flux drop associated with X--ray weak states is sufficient
  to depress the C~\textsc{iv} emission--line flux, or that the
  outflow can be launched from closer in during X--ray weak states, thus
  resulting in a larger outflow velocity and C~\textsc{iv}
  blueshift. The optical spectrum of PHL~1092 is dominated by
  Fe~\textsc{ii}, and is characterised by relatively narrow H$\beta$
  and Mg~\textsc{ii} lines centred at rest--frame energies, likely
  originating in the dense outer disc. No significant spectral changes
  are seen in the optical from 2008 to 2010.

\item The X--ray variability of PHL~1092 on long-- and
  short--timescales suggests a correlation between the X--ray flux and
  the RMS variability, consistent with the linear RMS--flux
  relationship that has been now observed in almost all types of
  accreting objects. PHL~1092 has a high variability amplitude on
  short--timescales which can be used to estimate its BH mass. The
  X--ray variability BH mass estimate is, however, about two orders of
  magnitude smaller than that obtained using the standard BH mass
  scaling relations using single--epoch optical spectra. This result calls
  into question either the idea that the mechanism behind the
  short-timescale variability of PHL~1092 is the same as in all other
  accreting objects (which would be surprising, given the RMS--flux
  relation we suggest) or the reliability of the standard mass scaling
  relations for wind--dominated {\it soft--spectrum} AGN such as
  PHL~1092.

\item We propose three possible scenarios that can explain the
  UV--to--X--ray SED of PHL~1092 at all X--ray flux levels: i) a {\it
    breathing corona} scenario in which the X--ray flux is correlated
  with the X--ray corona size and where X--ray weak states are
  associated with the collapse of the corona down to the marginal
  stable orbit; ii) an absorption model where the X--ray variability
  is dominated by changes of the covering fraction of a relatively
  cold absorber which covers almost 100\% of the X--ray source in
  X--ray weak states; iii) a disc--reflection--dominated scenario
  where the X--ray corona is confined within few $r_g$ from the
  central BH at all flux levels.

\end{enumerate}

In the {\it breathing corona} scenario, the X--ray corona of PHL~1092
is always compact ($\leq 2~R_{\rm{isco}}$) and shrinks down to $\sim
R_{\rm{isco}}$ in extreme X--ray weak states. However, the shrinking
radial velocity is likely too high to be plausible and, as such, we
interpret the {\it breathing corona} as an indication that the X--ray
active regions are more compact (and likely closer in) in low X--ray
flux states. If so, we caution that light--bending effects will
inevitably occur, so that the model reduces to the
reflection--dominated one at least in extreme X--ray weak states.

The absorption interpretation is not completely self--consistent as
the X--ray flux variability is produced by variations of a relatively
cold absorber. The absorber should imprint some absorption feature in
the UV, but UV absorption is not detected in the spectra of PHL~1092,
PHL~1811, the PHL~1811 analogs of Wu et al. (2011), or the two NLS~1
galaxies we consider. One possibility is that the absorber is confined
within the UV--emitting disc, i.e. within a few tens of $r_g$. If so,
however, absorbing clouds cannot survive at the low ionisation level
we need to reproduce the X--ray variability. A possible way out would
be to consider a multi--layer absorption model in which every
absorbing cloud is structured into multiple layers. The outer layers
may protect the innermost cloud(s) nucleus from the ionising flux,
thus enabling them to survive close to the source at low
ionisation. We defer such complex study to future work.

As mentioned, the X--ray variability and overall SED can also be
reproduced in a reflection-dominated scenario where a compact X--ray
corona is confined within few $r_g$ from the BH at all flux
levels. Light--bending effects produce a disc reflection--dominated
spectrum, and the observed extreme X--ray flux variability at soft
X--rays (a factor $260$ in 3.5~yr) is reproduced with a much smaller
factor of $\sim 10$ intrinsic variation. A similar scenario has been
invoked for 1H~0707--495 and IRAS~13224--3809, two NLS~1 galaxies
which share many properties with PHL~1092 from UV to X--rays. The
recent detection of soft X--ray time lags in 1H~0707--495 (and other
sources) supports a reflection scenario in these objects.

Two of the the three scenarios we present make very different
predictions for the X--ray flux above 10~keV. The absorption model
implies a constant flux $F_{30-100} \sim 2.5\times
10^{-13}$~erg~cm$^{-2}$~s$^{-1}$ in the 30--100~keV band, while the
reflection--dominated model predicts a much lower flux\footnote{We use
  as upper limit the hard X--ray flux predicted by the highest flux
  observation model, which is close to the historical highest X--ray
  flux of PHL~1092, see Fig.~\ref{xlc}.} of $F_{30-100} \leq 4.1\times
10^{-14}$~erg~cm$^{-2}$~s$^{-1}$ On the other hand, as the baseline
model is associated with very large intrinsic variability, the
predicted hard X--ray flux is too variable to provide a useful
constraint. A deep observation of PHL~1092 (or of a similar source)
with sensitive hard X--ray detectors such as {\it ASTRO--H} (Takahashi
et al. 2010) and {\it NuSTAR} (Harrison et al. 2010) in the near
future may thus reveal the nature of the transient X--ray weakness
phenomenon, and could shed light into the nature of the non--BAL
X--ray weak quasars population at large.

\section*{Acknowledgements}

Based on observations obtained with XMM-Newton, an ESA science mission
with instruments and contributions directly funded by ESA Member
States and NASA. We used observations made with the NASA/ESA Hubble
Space Telescope, obtained from the Data Archive at the Space Telescope
Science Institute, which is operated by the Association of
Universities for Research in Astronomy, Inc., under NASA contract
NAS~5-26555. The Hobby-Eberly Telescope (HET) is a joint project of
the University of Texas at Austin, the Pennsylvania State University,
Stanford University, Ludwig-Maximillians-Universit\"at M\"unchen, and
Georg-August-Universit\"at G\"ottingen.  The HET is named in honor of
its principal benefactors, William P.~Hobby and Robert E.~Eberly.  The
Marcario Low-Resolution Spectrograph is named for Mike Marcario of
High Lonesome Optics, who fabricated several optics for the instrument
but died before its completion; it is a joint project of the
Hobby-Eberly Telescope partnership and the Instituto de
Astronom\'{\i}a de la Universidad Nacional Aut\'onoma de
M\'exico. Funding for the SDSS and SDSS-II has been provided by the
Alfred P.~Sloan Foundation, the Participating Institutions, the
National Science Foundation, the U.S. Department of Energy, the
National Aeronautics and Space Administration, the Japanese
Monbukagakusho, the Max Planck Society, and the Higher Education
Funding Council for England.  The SDSS Web site \hbox{is {\tt
    http://www.sdss.org/}.} GM thanks the Spanish Ministerio de
Ciencia e Innovaci\'on and CSIC for support through a Ram\'on y Cajal
contract. Financial support for this work was also provided by the
Spanish Ministry of Science and Innovation through grant
AYA2010-21490-C02-02.  GM also thanks CSIC grant PA1003039 and the IoA
visitor program for support during an extended Summer visit in
Cambridge. WNB thanks NASA ADP grant NNX11AJ59G and NASA grant
NNX09AP83G for support. ACF thanks the Royal Society for support. GM
wishes to thank Fabrizio Nicastro and Enrico Piconcelli for useful
discussions.

\newpage 

\begin{table*}
  \caption{PHL~1092 X--ray observations, X--ray/UV flux densities, and
    UV/optical to X--ray spectral indices. T$_{\rm{exp}}$ is in units of
    ks and refers to the net exposure in the EPIC--pn detector after
    filtering for background periods. All observations were performed in
    the {\it XMM--Newton} EPIC Full Window (FW) observing mode except
    XMM(3), for which the Large Window (LW) mode was used. For XMM(1) no
    spectral information is however available and the flux has been
    determined from the EPIC--pn light curve. The 2~keV flux density
    f$_{\rm{2keV}}$ is given in units of
    $10^{-32}$~erg~s$^{-1}$~cm$^{-2}$~Hz$^{-1}$, and the UV/optical ones
    in units of $10^{-27}$~erg~s$^{-1}$~cm$^{-2}$~Hz$^{-1}$. The typical
    uncertainty on flux densities is 4--5 per cent at most (e.g. W2 filter). The rest--frame
    2500~\AA\ flux density is extrapolated from the the nearest
    available OM filter using a spectral index of $\alpha_\nu = -0.70$
    (Wu et al. 2001). $\Delta\alpha_{\rm{ox}}$ is the difference between
    the observed value and that expected from the 2500~\AA\ luminosity
    using the well known anti--correlation between $\alpha_{\rm{ox}}$
    and L$_{\rm{2500}}$ ($\alpha_{\rm{ox}}^{\rm{expected}} = -1.48$ for
    PHL~1092). The rest--frame effective wavelength of the OM filters
    are: 1519~\AA\ (W2); 1655~\AA\ (M2); 2084.5~\AA\ (W1);
    2464~\AA\ (U).}
\label{tab1}      
\begin{center}
\begin{tabular}{l c c c c c c c c c c c c}
\hline
\multicolumn{13}{l}{{\bf XMM--Newton observations and optical/UV/X--ray properties}} 
\\
\hline
Obs. & Date & T$_{\rm{exp}}$ & Obs. mode & f$_{\rm{2keV}}$ & f$_{\rm{W2}}$& f$_{\rm{M2}}$ & f$_{\rm{W1}}$ & f$_{\rm{U}}$ & f$_{\rm{2500}}$ & $\alpha_{\rm {W2x}}$ & $\alpha_{\rm{ox}}$ & $\Delta\alpha_{\rm{ox}}$\\ \\
XMM(1)$^*$ & 2000-07-31 & $25$ &FW& $10$   &$2.81$ &        &        &        & $3.98$ & $-1.86$ & $-1.77$ & $-0.29$\\
XMM(2)$^*$ & 2003-07-18 & $23$ &FW& $31$   &$2.69$ &        &        &        & $3.81$ & $-1.65$ & $-1.57$ & $-0.09$\\
XMM(3)$^*$ & 2008-01-20 & $57$ &LW& $0.12$ &$2.94$ &        &        &        & $4.17$ & $-2.67$ & $-2.51$ & $-1.03$\\
XMM(4)     & 2009-06-27 & $24$ &FW& $0.51$ &$2.52$ & $3.02$ & $3.23$ & $3.32$ & $3.35$ & $-2.38$ & $-2.23$ & $-0.75$\\
XMM(5)     & 2010-07-07 & $22$ &FW& $3.1$  &$2.72$ & $2.75$ & $2.99$ &        & $3.40$ & $-2.07$ & $-1.93$ & $-0.45$\\ 
XMM(6)     & 2010-12-30 & $25$ &FW& $2.4$  &$2.68$ & $2.88$ & $3.01$ & $3.20$ & $3.23$ & $-2.11$ & $-1.97$ & $-0.49$ \\
\hline                        
\end{tabular}
\end{center}
\raggedright 
$^*$ For the first three observations, f$_{\rm{2500}}$ is
obtained by extrapolating f$_{\rm{W2}}$ assuming a spectral index of
$\alpha_\nu = -0.70$. However, as discussed in Section~\ref{Cvar}, the W2 filter is
centred at 1519~\AA\, which coincides with the wing of  the broad, blueshifted, and
possibly variable C~\textsc{iv} emission line. Hence, f$_{\rm{2500}}$ is
not highly reliable, which affects the derived values of $\alpha_{\rm
  {W2x}}$, $\alpha_{\rm{ox}}$, and $\Delta\alpha_{\rm{ox}}$.  
\end{table*}


\begin{table*}
\caption{Best--fitting parameters for the phenomenological model
  comprising Galactic absorption, a multi--coloured blackbody, and a
  power--law (with slope forced to be the same in all the {\it
    XMM--Newton} observations). The inner disc blackbody temperature
  is not redshift--corrected and is given in units of keV. Fluxes in
  the 0.5--2~keV and 2--10~keV band are in c.g.s. units.}
\label{pheno}
\begin{center}
\begin{tabular}{l c c c c c}
\hline
\multicolumn{5}{l}{{\bf Phenomenological model }} & \multicolumn{1}{r}{$\chi^2/{\rm{dof}}=425/357$}
\\
\hline
Obs. & kT$_{\rm{in}}$  & N$_{\rm{diskbb}}$ & $\Gamma$ & $\log$~F$_{\rm{0.5-2}}$ & $\log$~F$_{\rm{2-10}}$ 
\\ \\ 
XMM(2) & $0.11\pm 0.01$ & $1520\pm 187$ & $2.0\pm 0.3$ &$-12.254\pm 0.007$ &$-12.76\pm 0.07$\\
XMM(3) & $0.08\pm 0.03$ & $504\pm 500$  & $-$          &$-14.630\pm 0.094$ &$\leq -14.20$\\
XMM(4) & $0.11\pm 0.03$ & $30\pm 25$    & $-$          &$-14.088\pm 0.063$ &$\leq -14.00$\\
XMM(5) & $0.10\pm 0.02$ & $1666\pm 298$ & $-$          &$-12.902\pm 0.014$ &$-13.63\pm 0.10$\\
XMM(6) & $0.10\pm 0.01$ & $373\pm 93$   & $-$          &$-13.190\pm 0.018$ &$-13.95\pm 0.10$\\
\hline 
\end{tabular}
\\
\end{center}
\end{table*}

\begin{table*}
\caption{Best--fitting parameters for the baseline model. Columns with
  only one entry refer to parameters that have been forced to be the
  same as in the previous row. We report results for both $a=0$ and
  $a=0.998$ as they are statistically equivalent. The black hole mass
  is in units of $10^8~M_\odot$. The outer corona radius R$_{\rm c}$
  is in units of $r_g = GM_{\rm{BH}}/c^2$ and the temperature of the
  electron population responsible for the soft X--ray excess
  ($kT_{\rm{e}}$) is in units of keV. The best--fitting X-ray unabsorbed
  luminosity is also reported for reference in units of
  $10^{43}$~erg~s$^{-1}$ in the soft 0.5--2~keV and hard 2--10~keV
  band for the $a=0$ model (as they are of course identical for the
  $a=0.998$ case). We do not report errors as these luminosities
  are mostly based upon model extrapolation.}
\label{mod1}      
\begin{center}
\begin{tabular}{l c c c c c c c c c}
\hline
\multicolumn{8}{l}{{\bf Baseline model ($a=0$)}} & \multicolumn{2}{r}{$\chi^2/{\rm{dof}}=480/370$}
\\
\hline
Obs. & M$_{\rm{BH}}$  &$\log$($L/L_{\rm{Edd}}$) & R$_{\rm c}$ & $kT_{\rm
  e}$ & $\tau_{\rm e}$ & $\Gamma_{\rm h}$ & $f_{\rm{pl}}$ & $L_{0.5-2}$ & $L_{2-10}$  
\\ \\ 
XMM(2) & $2.37\pm 0.25$ & $0.39\pm 0.08$ & $9.80\pm 0.70$ &$0.12\pm 0.01$& $42\pm 16$ & $2.8\pm 0.2$   & $0.29\pm 0.11$ & $65.0$ & $5.88$\\ 
XMM(3) & $-$ & $-$                       & $6.39\pm 0.08$ &$-$             & $-$      & $-$            & $0.55\pm 0.15$ & $0.29$ & $0.04$\\
XMM(4) & $-$ & $-$                       & $6.61\pm 0.08$ &$-$             & $-$      & $-$            & $0.45\pm 0.15$ & $0.98$ & $0.11$\\
XMM(5) & $-$ & $-$                       & $8.06\pm 0.32$ &$-$             & $-$      & $2.1\pm 0.3$   & $0.05\pm 0.04$ & $19.0$ & $0.77$\\ 
XMM(6) & $-$ & $-$                       & $7.48\pm 0.20$ &$-$             & $-$      & $-$            & $0.06\pm 0.04$ & $8.77$ & $0.57$\\
\hline 
\multicolumn{8}{l}{{\bf Baseline model ($a=0.998$)}} & \multicolumn{2}{r}{$\chi^2/{\rm{dof}}=480/370$}
\\
\hline
\\ 
XMM(2) & $17.0\pm 2.0$ & $-0.51\pm 0.08$ & $1.60\pm 0.06$ &$0.12\pm0.01$ & $42\pm 15$ & $2.8\pm 0.2$    & $0.30\pm 0.11$ & $-$ & $-$\\ 
XMM(3) & $-$ & $-$                       & $1.27\pm 0.01$ &$-$           & $-$        & $-$             & $0.56\pm 0.13$ & $-$ & $-$\\
XMM(4) & $-$ & $-$                       & $1.29\pm 0.01$ &$-$           & $-$        & $-$             & $0.43\pm 0.12$ & $-$ & $-$\\
XMM(5) & $-$ & $-$                       & $1.41\pm 0.04$ &$-$           & $-$        & $2.1\pm 0.3$    & $0.05\pm 0.04$ & $-$ & $-$\\ 
XMM(6) & $-$ & $-$                       & $1.36\pm 0.03$ &$-$           & $-$        & $-$             & $0.08\pm 0.05$ & $-$ & $-$\\
\hline 
\end{tabular}
\\
\end{center}
\end{table*}


\begin{table*}
\caption{Best--fitting parameters for the absorption model. The soft
  excess is modelled as Comptonization in an optically thick corona,
  and the intrinsic X--ray flux is assumed to be constant, i.e. one
  and the same in all the {\it XMM--Newton} observations. The coronal
  parameters are as follows: $kT_{\rm{e}} = 0.19\pm 0.2$~keV, $\tau =
  15\pm 8$, $\Gamma_{\rm{h}} = 1.8\pm 0.1$,  R$_{\rm c} = 15\pm 5~r_g$, and $f_{\rm{pl}} = 0.05\pm
  0.03$. All the above parameters are forced to be the same at all
  flux levels. The absorber column density is in units of
  $10^{22}$~cm$^{-2}$. Its ionisation state reaches the lower limit
  defined in the absorption model we use ($\log\xi = -3.0$), as
  indicated by the subscript $p$ (``pegged''). The
  absorption line energy is in units of keV and is given in the
  observed--frame.}
\label{mod2}      
\begin{center}
\begin{tabular}{l c c c c c c  }
\hline
\multicolumn{5}{l}{{\bf Absorption model ($a=0$)}} & \multicolumn{2}{r}{$\chi^2/{\rm{dof}}=410/355$}
\\ \hline 
\multicolumn{2}{l}{M$_{\rm{BH}} \equiv 2.37\times 10^8~M_\odot$}
& \multicolumn{5}{l}{$\log$(L/L$_{\rm{Edd}}$)$\equiv 0.39$} \\
\hline
Obs.   & $N_{\rm{H}}$ & $\log \xi$ & $C_f$ &$E_{\rm{abs}}$ & $\sigma_{\rm{abs}}$ & $\tau_{\rm{abs}}$ \\ \\
XMM(2) & $100_{\rm{fix}}$ & $-3.0^{+2.0}_{-0.0p}$  & $0.190\pm 0.010$ & $1.05\pm 0.03$ & $0.20\pm 0.02$ & $0.15\pm 0.03$ \\ 
XMM(3) & $-$            & $-$                & $0.996\pm 0.001$ & $-$            & $-$            & $\leq 0.60$    \\ 
XMM(4) & $-$            & $-$                & $0.985\pm 0.005$ & $-$            & $-$            & $\leq 0.60$    \\
XMM(5) & $-$            & $-$                & $0.725\pm 0.008$ & $-$            & $-$            & $1.00\pm 0.10$ \\
XMM(6) & $-$            & $-$                & $0.884\pm 0.005$ & $-$            & $-$            & $0.60\pm 0.10$ \\
\hline 
\end{tabular}
\\
\end{center}
\end{table*}


\begin{table*}
  \caption{Best--fitting parameters for the reflection--dominated
    model. The column density of the warm absorber (WA) is in units of
    $10^{22}$~cm$^{-2}$ and the disc inclination $i$ in
    degrees. $F_{\rm{ref}}$ is the total reflection flux over the whole
    band (here 0.001--300~keV) in units of
    $10^{-11}$~erg~s$^{-1}$~cm$^{-2}$. Notice that $F_{\rm{ref}}$ is
    likely over--estimated (but in the same way in all observations)
    because of the unknown low--energy bound which should be set at the
    energy below which the efficiency of the Compton up--scattering of
    the soft UV/EUV disc photons producing the irradiating power law is
    negligible.}
\label{mod34}      
\begin{center}
\begin{tabular}{l c c c c c c c c}
\hline
\multicolumn{7}{l}{{\bf Disc Reflection model ($a=0.998$)}} &  \multicolumn{2}{r}{$\chi^2/{\rm{dof}}= 420/353$}
\\ \hline 
\multicolumn{2}{l}{M$_{\rm{BH}} \equiv 17.0 \times 10^8~M_\odot$} & \multicolumn{7}{l}{$\log$(L/L$_{\rm{Edd}}$)$\equiv -0.51$} \\
\hline 
Obs. & N$_{\rm{H}}^{\rm(WA)}$ & $\log\xi^{\rm{(WA)}}$ & $q$ & $i$ & $\log\xi_{\rm{ref}}$ 
& $F_{\rm{ref}}$ & R$_{\rm c}$ & $\Gamma_{\rm h}$ \\ 
\\
XMM(2) & $0.05^{+0.03}_{-0.00p}$ & $-2.1\pm 0.2$      & $5.0\pm 0.3$ &$53\pm 6$ & $3.03\pm 0.04$ & $9.2\pm 0.4$  & $1.235_{\rm{fix}}$ & $2.57\pm 0.02$\\ 
XMM(3) & $-$                  & $-3.0^{+0.1}_{-0.0p}$ & $-$          &$-$       & $1.93\pm 0.17$ & $0.7\pm 0.3$  & $-$ & $-$\\ 
XMM(4) & $-$                  & $-3.0^{+0.1}_{-0.0p}$ & $-$          &$-$       & $2.45\pm 0.14$ & $2.6\pm 0.2$  & $-$ & $2.76\pm 0.03$\\ 
XMM(5) & $-$                  & $-2.5\pm 0.2$      & $-$          &$-$       & $2.74\pm 0.10$ & $4.5\pm 0.2$  & $-$ & $-$\\ 
XMM(6) & $-$                  & $-2.6\pm 0.2$      & $-$          &$-$       & $2.58\pm 0.10$ & $3.5\pm 0.2$  & $-$ & $-$\\ 
\hline
\end{tabular}
\\
\end{center}
\end{table*}

\label{lastpage}

\end{document}